\documentclass{article}
\usepackage{graphicx} 
\usepackage{amsmath,amssymb,amsthm}
\usepackage{colortbl}

\usepackage[margin=20truemm]{geometry}

\newtheorem{lemma}{Lemma}
\newtheorem{definition}{Definition}
\newtheorem{theorem}{Theorem}
\newtheorem{proposition}{Proposition}

\usepackage{url}
\usepackage{color}

\newcommand{\bra}[1]{\langle {#1} |}
\newcommand{\ket}[1]{| {#1} \rangle}

\newcommand{\ketbra}[1]{| {#1} \rangle\langle {#1} |}

\newcommand{\idop}{\mathbb{I}}

\newcommand{\tr}[1]{{\rm tr}\left[#1\right]}
\newcommand{\rank}[1]{{\rm rank}\left(#1\right)}
\newcommand{\range}[1]{{\rm range}\left(#1\right)}
\newcommand{\ptr}[2]{{\rm tr}_{#1}\left[#2\right]}
\newcommand{\lpnorm}[2]{\left\|#2\right\|_{#1}}
\newcommand{\lMnorm}[1]{\left\|#1\right\|_{\mathbb{M}}}
\newcommand{\conv}[1]{{\rm conv}\left(#1\right)}
\newcommand{\vspan}[1]{{\rm span}\left(#1\right)}

\newcommand{\pos}[1]{\mathbf{Pos}\left(#1\right)}
\newcommand{\dop}[1]{\mathbf{D}\left(#1\right)}

\newcommand{\cd}{\mathbb{C}^d}
\newcommand{\cdim}[1]{\mathbb{C}^{#1}}

\newcommand{\hh}{\mathcal{H}}
\newcommand{\vv}{\mathcal{V}}

\title{Duality of extremal quantum states in verification and data hiding
}
\author{Seiseki Akibue\thanks{Communication Science Laboratories, NTT, Inc., 3-1 Morinosato Wakamiya, Atsugi, Kanagawa 243-0198, Japan} and Yuki Takeuchi\thanks{Information Technology R\&D Center, Mitsubishi Electric Corporation, 5-1-1 Ofuna, Kamakura, Kanagawa 247-8501, Japan}}

\begin{document}

\maketitle
\abstract{Quantum state verification (QSV) and quantum data hiding (QDH) have so far been studied separately.
QSV decides whether a given quantum state is close to the ideal one, with significantly lower sample complexity compared with direct application of quantum tomography.
On the other hand, QDH is a cryptographic protocol that encodes secret classical information in multipartite quantum states, providing stronger security guarantees than conventional classical secret-sharing schemes.

Here, we consider two fundamental quantities of a pure quantum state, determining the sample complexity needed for QSV or the security level in QDH. We demonstrate that a pure state is most difficult to verify in QSV if and only if it is most secure in QDH with respect to those quantities. Furthermore, for such extremal states, the two fundamental quantities coincide.
We also generalize this relationship to one between the security of QDH using mixed states and the sample complexity for quantum subspace verification, which is a generalization of QSV.
As an application, we show the existence of efficient QSV protocols that are generalizations of the existing ones and examine the fundamental limitations on QDH by synthesizing research from both domains.}

\section{Introduction}

Quantum state verification (QSV)~\cite{PLM18,ZH19}, sometimes called state certification, decides whether an output state $\rho$ of a given quantum device is sufficiently close to a target state $|\Phi\rangle$; i.e., it distinguishes between cases where the fidelity $\langle\Phi|\rho|\Phi\rangle$ is high or low.
In addition to being of fundamental and practical importance of verifying quantum states, QSV has been applied to several quantum information processing tasks, such as demonstration of quantum advantage~\cite{TM18,CGKM21}, cloud quantum computation~\cite{HM15}, and quantum metrology~\cite{TMMSM19}, and proof-of-principle experiments have been conducted with small-scale photonic platforms~\cite{ZZCPXYYYHXCLG20,JWQCCLXSZM20}.
Here, let $\epsilon\in(0,1)$ and $\delta\in(0,1)$ be parameters specified by a verifier (i.e., an experimentalist verifying one's own quantum device).
Since non-orthogonal quantum states cannot be distinguished with certainty, the best we can do is to devise a QSV protocol such that the ideal $n$-qubit state $|\Phi\rangle$ can pass it with (almost) unit probability, but any noisy $n$-qubit state $\rho$ satisfying $\langle\Phi|\rho|\Phi\rangle\le1-\epsilon$ is rejected by it with probability at least $1-\delta$.
QSV protocols require multiple copies $\rho^{\otimes m}$ to achieve this goal, and the main theoretical objective in the field of QSV is to minimize $m$, which is called the {\it sample complexity}, for given $n$, $\epsilon$, and $\delta$.

It is known that if measurements of any kind are allowed in QSV protocols, any pure state $|\Phi\rangle$ can be verified with a sample complexity $O(\log(\delta^{-1})/\epsilon)$ that is independent of the number of qubits $n$.
More precisely, this sample complexity is optimal and is achieved by performing a positive operator-valued measure (POVM) measurement $\{|\Phi\rangle\langle\Phi|,\idop-|\Phi\rangle\langle\Phi|\}$ and accepting the given state $\rho$ only if the measurement outcome $\ketbra{\Phi}$ is obtained consistently across all $m$ copies of the sample.
However, accurate implementation of this POVM would be, in general, highly burdensome because we have to consider situations where accurate generation of $|\Phi\rangle$ is not easy and hence we would like to verify it.

Another approach, quantum state tomography, enables QSV to be performed with only single-qubit measurements by reconstructing a matrix representation of $\rho$.
However, the sample complexity is $\Omega(4^n)$ for constant $\epsilon$ and $\delta$.
Given that the quantum state tomography estimates not only the value of the fidelity but also a complete description of the target state, it is expected that we can construct much more efficient QSV protocols even with restricted measurements.

Indeed, for some quantum states such as graph states and their generalizations hypergraph and weighted graph states, the optimal sample complexity $O(\log(\delta^{-1})/\epsilon)$ is achieved with only single-qubit measurements  (see also Table I in Ref.~\cite{ZH19A}).
On the other hand, a known QSV protocol requires $\Omega(n\log(\delta^{-1})/\epsilon)$ samples of Dicke states~\cite{LYSZZ19}.
It was also shown that $O(n^2\log(\delta^{-1})/\epsilon)$ samples are sufficient to verify typical random pure states with single-qubit measurements~\cite{HPS24}. 
Numerical studies suggest that the sample complexity can be independent of the number $n$ of qubits even when measurements are restricted to being non-adaptive single-qubit ones~\cite{ATM24,LZ25}. Very recently, it has been shown that any pure state can be verified using adaptive single-qubit measurements with $O(n\log(\delta^{-1})/\epsilon)$ samples \cite{GHO25}. In that work, the authors also demonstrated the existence of a target state whose sample complexity is exponential in $n$ when only non-adaptive measurements are permitted. However, establishing nontrivial lower bounds on the sample complexity for general target states has remained a challenging open problem
because determination of the sample complexity requires optimization of local measurements, which is, in general, computationally intractable.
In fact, the existing analyses of sample complexity rely on the structure or randomness of the target states~\cite{HPS24,GHO25} or are restricted to a small number of qubits~\cite{PLM18,TDS20}.

\subsection{Our contribution}
\begin{figure}[ht]
    \centering
    \includegraphics[width=18cm]{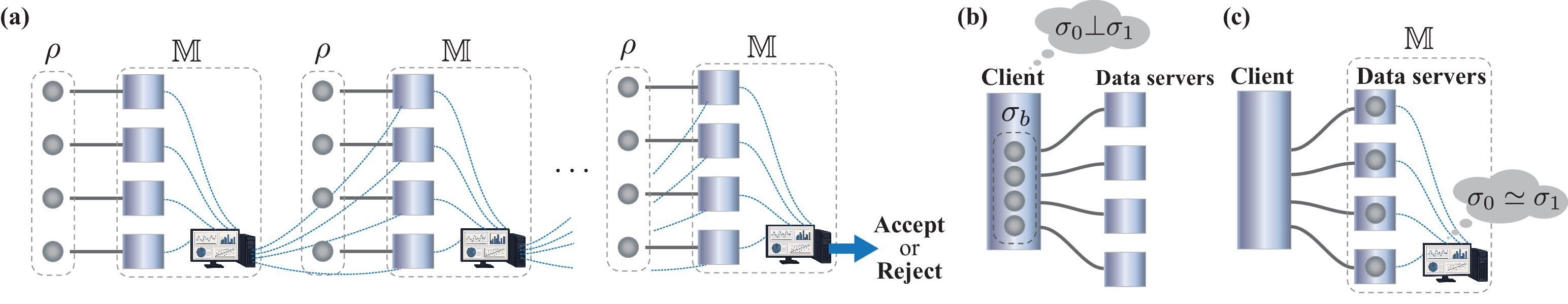}
    \caption{ Illustration of QSV and QDH protocols. A measurement class $\mathbb{M}$, which the verifier in QSV and the data servers in QDH are permitted to use, is typically assumed to be the set of LOCC measurements. (a) A QSV protocol based on a sequential measurement. For each sample of $\rho$, a measurement in $\mathbb{M}$ is performed that can depend on the previous measurement outcomes. However, entangled measurements across multiple samples of $\rho$ are not allowed. (b) The client encodes a secret bit $b$ into a quantum state $\sigma_b$ such that it can be perfectly recovered through an appropriate measurement. (c) Data servers, which are only able to perform measurements in $\mathbb{M}$, are unable to identify the encoded secret bit $b$.}
    \label{fig:QSVQDH}
\end{figure}

In this paper, we describe a way to tackle the above problem by connecting QSV to quantum data hiding (QDH)~\cite{TDL01,DLT02}.
QDH is a classically unrealizable cryptographic protocol, that works as follows.
First, a client generates an $n$-partite state $\sigma_b$ that encodes a secret single bit $b\in\{0,1\}$.
Then, the client sends each part of $\sigma_b$ to each of $n$ distant data servers.
QDH guarantees that the client can recover the original bit $b$ from $\sigma_b$ with an appropriate global measurement (see Fig.~\ref{fig:QSVQDH}  (b)),
while the servers cannot eavesdrop on the value of $b$ even if they perform local operations and classical communication (LOCC) (see Fig.~\ref{fig:QSVQDH}  (c)).
Therefore, we can quantify the security of QDH in terms of the indistinguishability of $\sigma_0$ and $\sigma_1$ when using local measurements.

We demonstrate an equivalence between the hardness of QSV under a measurement class $\mathbb{M}$ and the security of QDH against malicious data servers capable of performing measurements within $\mathbb{M}$ in the following sense.
\vspace{2.5mm}

\noindent \textbf{Duality between QSV and QDH (special case):}
\begin{itemize}
 \item For a pure state $\ket{\Phi}$, if there does not exist a secure QDH protocol using $\ket{\Phi}$ against a measurement class $\mathbb{M}$, there exists an efficient QSV protocol for verifying $\ket{\Phi}$ (with low sample complexity).

 \item For a pure state $\ket{\Phi}$, if there does not exist an efficient QSV protocol that uses $\mathbb{M}$ to verify the pure state $\ket{\Phi}$, then we can construct a set of states $\{\sigma_b\}_{b\in\{0,1\}}$, based on $\ket{\Phi}$, that is useful for QDH, where $\sigma_0$ is a pure state.

 \item We define two fundamental quantities $\gamma_{\Phi,\mathbb{M}}(\epsilon)$ and $\mu_{\Phi,\mathbb{M}}(\epsilon)$ of a pure state $\ket{\Phi}$. The sample complexity needed for QSV with sequential measurements performed on each sample of $\rho$ (see Fig.~\ref{fig:QSVQDH} (a)) is shown to be bounded between $\Omega((\gamma_{\Phi,\mathbb{M}}(\epsilon)\epsilon)^{-1}\log\frac{1}{\delta})$ and $O((\gamma_{\Phi,\mathbb{M}}(\epsilon)\epsilon)^{-2}\log\frac{1}{\delta})$. In the QDH scenario, there exists a quantum state $\sigma$ such that the probability of successfully identifying a randomly chosen state from $\{\Phi,\sigma\}$ is at least $\frac{1}{2}(1+\epsilon)$ whereas, under $\mathbb{M}$, the probability is at most $\frac{1}{2}(1+\mu_{\Phi,\mathbb{M}}(\epsilon)\epsilon)$.
 We show that a pure state $\ket{\Phi}$ minimizes $\gamma_{\Phi,\mathbb{M}}(\epsilon)$ if and only if it minimizes $\mu_{\Phi,\mathbb{M}}(\epsilon)$. Furthermore, for such extremal states $\ket{\hat{\Phi}}$ and for any $\epsilon$, both $\gamma_{\hat{\Phi},\mathbb{M}}(\epsilon)$ and $\mu_{\hat{\Phi},\mathbb{M}}(\epsilon)$ are equal to a universal constant, which is determined by $\mathbb{M}$.
 \end{itemize}

We show that this duality holds for any measurement class $\mathbb{M}$ that is informationally complete, which contains almost all the reasonable restricted classes of measurements, including LOCC, Pauli, and stabilizer measurements.
This duality opens up a new avenue for exploring QSV protocols by utilizing findings from QDH and vice versa. Additionally, it underscores the significance of rigorously proving sample complexity, not just for theoretical foundations, but also for the development of cryptographic protocols.

Although QSV and QDH are inherently relevant to state discrimination under restricted measurements, the duality we reveal is nontrivial.
This is because QSV only focuses on distinguishability under restricted measurements, but QDH is also related to distinguishability under unrestricted measurements.
Additionally, the measurement requirement in QSV is more stringent than in QDH because the former considers distinguishability of one state (target state $\ket{\Phi}$) among many states (low-fidelity states $\sigma$ such that $\bra{\Phi}\sigma\ket{\Phi}\leq1-\epsilon$) while the latter only considers distinguishability between two states.

Furthermore, we prove a more general form of the above duality: we show an equivalence between the security of QDH using mixed states and the hardness of quantum {\it subspace} verification, which is a generalization of QSV.
The task of quantum subspace verification is to determine whether the range of a given state $\rho$ is in a desired subspace $\vv$ or is far from it, i.e., $\tr{\rho\Pi_\vv}\leq1-\epsilon$, where $\Pi_\vv$ is the projector onto $\vv$ \cite{FH17}. From here on, we will refer to quantum subspace verification simply as QSV since quantum state verification can be considered a special instance of $\dim\vv=1$.
The duality can be summarized as follows.

\vspace{2.5mm}

\noindent \textbf{Duality between QSV and QDH (general case):}
\begin{itemize}
  \item For a subspace $\vv$, if there does not exist a secure QDH protocol using any $\rho$ lying in $\vv$ against a measurement class $\mathbb{M}$, there exists an efficient QSV protocol for verifying $\vv$ (with low sample complexity). (Theorem \ref{thm:comparison})

 \item For a subspace $\vv$, if there does not exist an efficient QSV protocol using $\mathbb{M}$ to verify the subspace $\vv$, then we can construct a set of states $\{\sigma_b\}_{b\in\{0,1\}}$, based on $\vv$, that is useful for QDH, where $\sigma_0$ is a quantum state whose rank is at most $\dim\vv$. (Theorem \ref{thm:constructionQDHfromQSV})
 
 \item We define two fundamental quantities $\gamma_{\vv,\mathbb{M}}(\epsilon)$ and $\mu_{\rho,\mathbb{M}}(\epsilon)$ of a subspace $\vv$ and a quantum state $\rho$. The sample complexity needed for QSV with sequential measurements performed on each sample of $\rho$ (see Fig.~\ref{fig:QSVQDH} (a)) is shown to be bounded between $\Omega((\gamma_{\vv,\mathbb{M}}(\epsilon)\epsilon)^{-1}\log\frac{1}{\delta})$ and $O((\gamma_{\vv,\mathbb{M}}(\epsilon)\epsilon)^{-2}\log\frac{1}{\delta})$. In the QDH scenario, there exists a quantum state $\sigma$ such that the probability of successfully identifying a randomly chosen state from $\{\rho,\sigma\}$ is at least $\frac{1}{2}(1+\epsilon)$ and that under $\mathbb{M}$, it is at most $\frac{1}{2}(1+\mu_{\rho,\mathbb{M}}(\epsilon)\epsilon)$.
 We demonstrate a bidirectional relationship between the extremal subspace and the state: (i) If a subspace $\vv$ minimizes $\gamma_{\vv,\mathbb{M}}(\epsilon)$ over the domain of subspaces whose dimension is at most $r$, there exists a state $\rho$ lying in $\vv$ such that $\rho$ minimizes $\mu_{\rho,\mathbb{M}}(\epsilon)$ over the domain of quantum states whose rank is at most $r$. (ii) If $\rho$ minimizes $\mu_{\rho,\mathbb{M}}(\epsilon)$ over the domain of quantum states whose rank is at most $r$, the subspace $\vv=\range{\rho}$ minimizes $\gamma_{\vv,\mathbb{M}}(\epsilon)$ over the domain of subspaces whose dimension is at most $r$.
 Furthermore, for such an extremal subspace $\hat{\vv}$ and state $\hat{\rho}$ and for any $\epsilon$, both $\gamma_{\hat{\vv},\mathbb{M}}(\epsilon)$ and $\mu_{\hat{\rho},\mathbb{M}}(\epsilon)$ are equal to a universal constant, which is determined by $\mathbb{M}$ and natural number $r$. (Theorem \ref{thm:equivalence})
 \end{itemize}

\section*{Application}
\subsection*{Implications from QDH to QSV}
Table \ref{table:implications} summarizes the implications arising from the duality between QSV and QDH, together with insights from related research on QDH. In what follows, we detail how these findings extend previously known results and discuss their broader significance.

\begin{table}
    \centering
    \begin{tabular}{ccc}
       \hline
       measurement class $\mathbb{M}$ & subspace $\vv$ & $\gamma_{\vv,\mathbb{M}}(\epsilon)$\\
       \hline
       (1) Non-adaptive single-qudit measurements on $(\cdim{d})^{\otimes n}$  & $\vee_n\cdim{d}$ & $\geq\frac{d}{\dim\vv}$\\
       (2) Non-adaptive single-qudit measurements on $(\cd)^{\otimes n}$  & arbitrary & $\geq\frac{1}{2\sqrt{18^n\dim\vv}}$\\
       (3) PPT measurements on $(\cd)^{\otimes n}$ & arbitrary & $\geq\frac{1}{2\sqrt{\dim\vv}}$\\
        \rowcolor[rgb]{0.8, 0.8, 0.8}
       (4) PPT measurements on $(\cd)^{\otimes n}$ & $\vee_n\cd$ & $\leq\frac{2}{d}$ ($n=2$), $\leq\frac{6n^2}{\sqrt{d}}$ ($n\geq3$)\\
       (5) $4$-design POVMs on $\cd$ & arbitrary & $\geq\frac{1}{6\sqrt{\dim\vv}}$\\
       \hline
    \end{tabular}
    \caption{New lower (non-shaded row) and upper bounds (shaded row) on the fundamental quantities $\gamma_{\vv,\mathbb{M}}(\epsilon)$ for verifying each subspace $\vv$ by using measurement class $\mathbb{M}$. $\vee_n\cd$ represents the symmetric subspace in $n$ qudits. Note that the lower and upper bounds on $\gamma_{\vv,\mathbb{M}}(\epsilon)$ respectively provide an upper bound $O((\gamma_{\vv,\mathbb{M}}(\epsilon)\epsilon)^{-2}\log\frac{1}{\delta})$ and a lower bound $\Omega((\gamma_{\vv,\mathbb{M}}(\epsilon)\epsilon)^{-1}\log\frac{1}{\delta})$ on the sample complexity of QSV.}
    \label{table:implications}
\end{table}

\begin{enumerate}
    \item We construct a QSV protocol for verifying an $n$-qudit symmetric subspace by using non-adaptive single-qudit measurements that has a sample complexity of $O((\dim\vee_n\cd)^2\log(\delta^{-1})/(\epsilon d)^2)$. The construction employs a symmetrization technique commonly used in the design of QDH protocols~\cite{DLT02,EW02,H23}. As a corollary, we obtain a QSV protocol for verifying any multi-qubit subspace spanned by a subset of Dicke states that has a sample complexity of $O(n^2\log(\delta^{-1})/\epsilon^2)$, thereby generalizing the result of Zheng et al. \cite{ZXWZ25} from the finite regime ($n\leq3$) to the asymptotic regime.

    \item We demonstrate the existence of a QSV protocol for verifying any $n$-qudit subspace $\vv$ by using non-adaptive single-qudit measurements that has a sample complexity of $O(18^n(\dim\vv)\log(\delta^{-1})/\epsilon^2)$, by leveraging a result of Lancien \emph{et~al.}~\cite{LW13}. This protocol generalizes Nathanson's protocol~\cite{N10}, which assumes $n = 2$ and $\dim \vv = 1$.

    \item We demonstrate the existence of a QSV protocol for verifying any $n$-qudit subspace $\vv$ by using positive partial transpose (PPT) measurements that has a sample complexity of $O((\dim\vv)\log(\delta^{-1})/\epsilon^2)$ by leveraging a result of Lancien \emph{et~al.}~\cite{LW13}. Here, the PPT measurement serves as a mathematical model for approximating a local measurement, and it is frequently used to prove the security of QDH. As a consequence, any multipartite pure state can be verified with a constant sample complexity, independent of both the number $n$ of qudits and the local dimension $d$ of each qudit. In contrast, achieving such a constant bound has remained a major open problem in the case of adaptive single-qudit measurements~\cite{HPS24,GHO25}. Furthermore, we construct a QSV protocol for verifying multipartite pure states and determine the universal constant $\min_\Phi\gamma_{\Phi,\mathbb{PPT}_n}(\epsilon)$ appearing in the duality.

    \item By leveraging the result of Harrow~\cite{H23}, we establish lower bounds of $\Omega(d\log(\delta^{-1})/\epsilon)$ for $n = 2$ and $\Omega(\sqrt{d}\log(\delta^{-1})/(n^2\epsilon))$ for $n \geq 3$ on the sample complexity of verifying the symmetric subspace by using PPT measurements.

    \item By leveraging a result of Matthews \emph{et~al.}~\cite{MSA09}, we demonstrate the existence of a QSV protocol for verifying any subspace using $4$-design POVMs that has a sample complexity of $O((\dim\vv)\log(\delta^{-1})/\epsilon^2)$. While the same sample complexity can be attained via shadow tomography~\cite{HKP20}, our protocol substantially simplifies the post-processing of the measurement outcomes. Specifically, in our protocol, it suffices to store only a single bit of information for each measurement applied to each sample of $\rho$, whereas classical shadow tomography requires $n$ bits to be stored and more involved post-processing to reconstruct a classical description of $\rho$.

\end{enumerate}

It is worth noting that we can derive lower bounds on $\gamma_{\vv,\mathbb{M}}(\epsilon)$ as a direct consequence of the duality between QSV and QDH and existence of the following type of inequality:
\begin{equation}
    \label{eq:distratio}
    \lpnorm{\mathbb{M}}{\rho-\sigma}\geq r\lpnorm{1}{\rho-\sigma},
\end{equation}
where the two norms $\lpnorm{\mathbb{M}}{\Delta}$ and $\lpnorm{1}{\Delta}$ represent distinguishability under a measurement class $\mathbb{M}$ and under unrestricted measurements, respectively. This type of inequality has been extensively studied from the fundamental point of view \cite{MSA09,LW13,KZG16,LPW18,CLP22,ZLZW24}.
Note that this approach to deriving lower bounds is totally different from the existing one where lower bounds are obtained by explicitly constructing QSV protocols.

\subsection*{Implications of QSV on QDH}

The duality has the following implications on the limitations and possibility of QDH using multiqubit pure states.

\begin{itemize}
    \item For any $n$-qubit pure state $\ket{\Phi}$, the fundamental quantity $\mu_{\Phi,\mathbb{LOCC}_n}(\epsilon)$ for adaptive single-qubit measurements is lower bounded by $\frac{1}{n}$. In other words, for any multiqubit pure state $\ket{\Phi}$ and any mixed state $\sigma$, 
    \begin{equation}
            \lpnorm{\mathbb{LOCC}_n}{\Phi-\sigma}\geq\frac{1}{n}\lpnorm{1}{\Phi-\sigma}
    \end{equation}
    holds, where $\mathbb{LOCC}_n$ is the class of adaptive single-qubit measurements.

    \item There exists a sequence of $n$-qubit pure states $\ket{\hat{\Phi}_n}$ for which the fundamental quantity $\mu_{\hat{\Phi}_n,\mathbb{LPV}_n}(1)$ for non-adaptive single-qubit projective measurements $\mathbb{LPV}_n$ is upper bounded as $2^{-\Omega(n)}$. In other words, there exist a sequence of $n$-qubit pure states $\ket{\hat{\Phi}_n}$ and mixed states $\hat{\sigma}_n$ for every $n$ such that $\hat{\Phi}_n$ and $\hat{\sigma}_n$ are orthogonal and
    \begin{equation}
            \lpnorm{\mathbb{LPV}_n}{\hat{\Phi}_n-\hat{\sigma}_n}= 2^{-\Omega(n)}
    \end{equation}
    holds.
\end{itemize}

The first implication establishes that any QDH protocol employing a pure state fails to achieve strong security against data servers capable of performing adaptive single-qubit measurements. While this aligns with expectations from the bipartite setting, a rigorous proof follows from incorporation of a recent breakthrough in QSV~\cite{GHO25} into our duality framework.

The second implication establishes that a secure QDH protocol employing a pure state is achievable when data servers are restricted to non-adaptive single-qubit projective measurements. Since $\mu_{\Phi,\mathbb{LO}}(1)$ is lower bounded by a constant~\cite{MSA09} independent of the dimension in the case of bipartite non-adaptive single-qudit measurements, $\mathbb{LO}$, and the ratio is still lower bounded by a constant even if the measurements are restricted to being projective~\cite{KO25}, this result reveals a sharp contrast in local distinguishability between the bipartite and multipartite scenarios. The recent breakthrough in QSV~\cite{GHO25} can also be used to derive this implication within our duality framework.

\section{Notation}
Let us briefly introduce the notation and concepts of quantum information in this subsection. Readers can find a more comprehensive introduction to quantum information in \cite{WBook, MHBook}.

We only consider finite-dimensional Hilbert spaces.
A pure state is represented by a unit vector $\ket{\phi}\in\hh$ in a Hilbert space $\hh$. Its density operator, denoted by $\phi:=\ketbra{\phi}$, is also often referred to as a pure state. A vector and a corresponding rank-1 operator that are not necessarily normalized are denoted by $\ket{\tilde{\phi}}$ and $\tilde{\phi}:=\ketbra{\tilde{\phi}}$, respectively.
For a linear operator $A:\hh\rightarrow\hh'$, its range is defined by $\range{A}:=\{A\ket{\tilde{\phi}}:\ket{\tilde{\phi}}\in\hh\}$.
$\pos{\hh}$ represents the set of positive semi-definite operators acting on a Hilbert space $\hh$.
$\idop\in\pos{\hh}$ represents the identity operator.
$\dop{\hh}$ represents the set of density operators $\rho$ that satisfy $\rho\in\pos{\hh}$ and $\tr{\rho}=1$.
For a finite set $I$, a POVM is a set $\{M_i\in\pos{\hh}\}_{i\in I}$ of positive semi-definite operators that is a resolution of unity; i.e., $\sum_{i\in I}M_i=\idop$.
$\Pi_\vv$ represents a Hermitian projector onto a subspace $\vv\subseteq\hh$.
All the logarithms used in this paper are the natural logarithms.

\section{Distinguishability measure under restricted measurements}
Since both QSV and QDH address different aspects of quantum state distinguishability, we begin by reviewing fundamental concepts related to quantum state discrimination. Specifically, we will define these concepts in scenarios where only specific types of restricted measurements can be performed, as both the hardness of QSV and the security of QDH arise in these contexts. 
First, we formally define $\mathbb{M}$ as the class (or set) of measurements that may have restrictions.
Mathematically, this class is merely a (not necessarily proper) subset of the set of POVMs.
If $\mathbb{M}$ is the set of POVMs, we refer to it as the unrestricted measurement class. We sometimes refer $\mathbb{M}$ as a restricted measurement class if it is a proper subset.

\begin{definition}
   For a Hilbert space $\hh$, a measurement class $\mathbb{M}$ is a subset of the set of POVMs over $\hh$. Note that $\mathbb{M}$ may contain POVMs that have a different number of outcomes.
\end{definition}

Next, we define the set of binary measurements related to $\mathbb{M}$, which is crucial for examining distinguishability.

\begin{definition}
\label{def:binaryM}
    We define the set $\mathbf{M}$ of binary measurements associated with $\mathbb{M}$ as
    \begin{equation}
    	 \mathbf{M}:=\left\{\sum_{j\in J}p(j)\sum_{i\in I_j}r(i,j)M_i^{(j)}:|J|<\infty, p(j)\geq0,\sum_{j\in J}p(j)=1, r(i,j)\in[0,1],\{M_i^{(j)}\}_{i\in I_j}\in \mathbb{M}\right\}.
 	\end{equation}
\end{definition}
Note that for any $M\in\mathbf{M}$, a POVM $\{M,\idop-M\}$ is realizable by using classical pre- and post-processing. We can also verify that the following properties of $\mathbf{M}$ hold.

\begin{itemize}
 \item (symmetry around $\frac{\idop}{2}$) $\idop-M\in\mathbf{M}$ if $M\in\mathbf{M}$,
 \item (trivial element) $0\in\mathbf{M}$, and
 \item (convexity) $\mathbf{M}$ is convex.
\end{itemize}

\begin{definition}
For a Hermitian operator $\Delta$ and a measurement class $\mathbb{M}$, the $\mathbb{M}$-norm is defined as
\begin{eqnarray}
        \lMnorm{\Delta}&:=&\sup_{\{M_i\}_{i\in I}\in \mathbb{M}}\sum_{i\in I}\left|\tr{M_i\Delta}\right|\\
        &=&\sup_{M\in \mathbf{M}}\left(\tr{M\Delta}-\tr{(\idop-M)\Delta}\right)\\
        &=&2\sup_{M\in \mathbf{M}}\tr{M\Delta}-\tr{\Delta}.
\end{eqnarray}
\end{definition}
We can verify that the $\mathbb{M}$-norm is a seminorm, i.e., that it satisfies the triangle inequality and absolute homogeneity.
Moreover, it is a norm if $\mathbb{M}$ is informationally complete; i.e., the real span of $\mathbf{M}$ is equal to the set of Hermitian operators on $\hh$.
If $\mathbb{M}$ is the unrestricted measurement class, i.e., $\mathbf{M}=\{M\in\pos{\hh}:M\leq\idop\}$, the $\mathbb{M}$-norm coincides with the Schatten 1-norm $\lpnorm{1}{\Delta}$.

This definition of the norm is essentially the same as the ones defined in \cite{MSA09}.
It is known that the $\mathbb{M}$-norm characterizes the distinguishability of two quantum states $\rho$ and $\sigma$ under $\mathbb{M}$.
For example, the maximal success probability $p_{succ}$ for distinguishing the two states when they are prepared with probability $\frac{1}{2}$ is given by $p_{succ}=\frac{1}{2}+\frac{1}{4}\lMnorm{\rho-\sigma}$.

It is known that two states $\rho$ and $\sigma$ are perfectly distinguishable under the unrestricted measurement class, i.e., $p_{succ}=1\Leftrightarrow\lpnorm{1}{\rho-\sigma}=2$, if and only if they are orthogonal, i.e., $\range{\rho}$ and $\range{\sigma}$ are orthogonal subspaces.
However, under a restricted measurement class, $\rho$ and $\sigma$ may not be perfectly distinguishable, i.e., $\lMnorm{\rho-\sigma}<2$, even if they are orthogonal.
Since the gap in distinguishability between the case of unrestricted and restricted measurement classes indicates the potential of QDH, the following quantity has been extensively analyzed \cite{KZG16,LPW18,CLP22,ZLZW24}.
\begin{definition}
\label{def:distratio}
 For a state $\rho\in\dop{\hh}$ and a measurement class $\mathbb{M}$, a universal constant $\hat{\mu}_{\rho,\mathbb{M}}\in[0,1]$, which we call the distinguishability ratio, is defined as follows.
 \begin{equation}
 	\hat{\mu}_{\rho,\mathbb{M}}:=\inf_{\sigma\in\dop{\hh}\setminus\{\rho\}}  \frac{\lMnorm{\rho-\sigma}}{\lpnorm{1}{\rho-\sigma}}.
\end{equation}
\end{definition}

Note that $\hat{\mu}_{\rho,\mathbb{M}}$ is small if and only if there exists a state $\sigma$ such that the gap of the distinguishability between $\rho$ and $\sigma$ under the unrestricted and restricted measurement classes is large. Since such $\rho$ and $\sigma$ are not necessarily orthogonal, they are not directly useful in QDH.

For example, let us consider a measurement class $\mathbb{LO}$ corresponding to the measurements realized by bipartite and non-adaptive local operations and classical communications (LOCC). Formally, $\mathbb{LO}$ is defined by
\begin{equation}
 \mathbb{LO}:=\left\{\left\{A_i\otimes B_j\right\}_{i,j}:\{A_i\}_{i\in I}\ and\ \{B_j\}_{j\in J}\ are\ POVMs\right\}.
\end{equation}
Then, the associated set $\mathbf{LO}$ of binary measurements is given by
\begin{equation}
 \mathbf{LO}:=\conv{\left\{\sum_{(i,j)\in R}A_i\otimes B_j:\{A_i\}_{i\in I}\ and\ \{B_j\}_{j\in J}\ are\ POVMs,R\subseteq I\times J\right\}}.
\end{equation}
It was shown \cite[Theorem 15]{MSA09} that
\begin{equation}
 \lpnorm{\mathbb{LO}}{\rho-\sigma}\geq\frac{1}{\sqrt{153}}\lpnorm{2}{\rho-\sigma}.
\end{equation}
We can verify this by observing that the closure of $\mathbf{LO}$ includes the set of binary measurements associated with the uniform POVM, which is used in \cite{MSA09} and has infinite measurement outcomes. 
Since $\lpnorm{2}{\rho-\sigma}=\sqrt{\tr{P^2+Q^2}}\geq\sqrt{\tr{P^2}}\geq\frac{\tr{P}}{\sqrt{\rank{P}}}=\frac{\lpnorm{1}{\rho-\sigma}}{2\sqrt{\rank{P}}}\geq\frac{\lpnorm{1}{\rho-\sigma}}{2\sqrt{\rank{\rho}}}$ where $P,Q\in\pos{\hh_A\otimes\hh_B}$ satisfy $P-Q=\rho-\sigma$ and $\tr{PQ}=0$, we obtain 
\begin{equation}
\label{eq:2LOratio}
 \hat{\mu}_{\rho,\mathbb{LO}}\geq\frac{1}{2\sqrt{153\rank{\rho}}}.
\end{equation}
This implies that realizing secure QDH requires a highly mixed state $\rho$ and the dimension of local systems must be large \cite{MSA09} .

\section{Quantum Data Hiding (QDH)}
Quantum data hiding (QDH) basically uses two nearly orthogonal quantum states $\sigma_0$ and $\sigma_1$ that are nearly indistinguishable under local measurements \cite{DLT02,EW02,HLSW04,AL14}.
First, we formally define the notion of data hiding for a pair of states.
\begin{definition}
 For $\epsilon,\delta\in(0,1]$, a pair of quantum states $(\sigma_0,\sigma_1)$ with $\sigma_i\in\dop{\hh}$ is called $(\epsilon,\delta)$-data hiding against measurement class $\mathbb{M}$ if it satisfies
 \begin{equation}
 \frac{1}{2}\lpnorm{1}{\sigma_0-\sigma_1}\geq\epsilon, \ \ \frac{1}{2}\lMnorm{\sigma_0-\sigma_1}\leq\delta.
\end{equation}
In other words, the success probability of identifying a randomly chosen state from $\{\sigma_0,\sigma_1\}$ can be at least $\frac{1}{2}(1+\epsilon)$, and that of identification under $\mathbb{M}$ is at most $\frac{1}{2}(1+\delta)$.
\end{definition}
By definition, 1-bit classical information $b\in\{0,1\}$ encoded in an $(\epsilon,\delta)$-data hiding pair $(\sigma_0,\sigma_1)$ is nearly undecodable under $\mathbb{M}$ if $\delta$ is small while it is almost perfectly decodable under the unrestricted measurement class if $\epsilon$ is large.
In this paper, we do not investigate QDH for quantum information \cite{HLSW04}; instead, but we consider another generalization of QDH in which $\mathbb{M}$ is not necessarily limited to the class of local measurements.

We characterize the security of QDH on the basis of the following quantity.
\begin{definition}
For a quantum state $\rho\in\dop{\hh}$, a measurement class $\mathbb{M}$ and a positive real $\epsilon\in(0,\epsilon_\rho]$, we define the $\epsilon$-distinguishability ratio $\mu_{\rho,\mathbb{M}}(\epsilon)\in[0,1]$ under $\mathbb{M}$ as follows:
\begin{equation}
\label{eq:defdistratio}
    \mu_{\rho,\mathbb{M}}(\epsilon):=\frac{1}{2\epsilon}\min_{\substack{\sigma\in\dop{\hh}:\lpnorm{1}{\rho-\sigma}\geq2\epsilon}}\lMnorm{\rho-\sigma},
\end{equation}
where $\epsilon_\rho=\frac{1}{2}\max_{\sigma\in\dop{\hh}}\lpnorm{1}{\rho-\sigma}$.
\end{definition}
Note that we can take the minimum (rather than infimum) due to the continuity of the semi-norm and the compactness of the region of $\sigma$.
By definition, for a state $\rho$, there exists a counterpart $\sigma$ such that $(\rho,\sigma)$ is $(\epsilon,\mu_{\rho,\mathbb{M}}(\epsilon)\epsilon)$-data hiding, and there does not exist such a pair of quantum states including $\rho$ that is $(\epsilon,\delta)$-data hiding if $\delta<\mu_{\rho,\mathbb{M}}(\epsilon)\epsilon$.
Thus, we find that a smaller value of $\mu_{\rho,\mathbb{M}}(\epsilon)$ enables the design of more secure QDH protocols.

The case of $\epsilon=1$ corresponds to an important class of QDH where a secret bit $b$ is perfectly decodable \cite{DLT02,EW02}. The expression of $\mu_{\rho,\mathbb{M}}(1)$ can be simplified as
\begin{equation}
\label{eq:distratio1}
    \mu_{\rho,\mathbb{M}}(1)=\frac{1}{2}\min_{\substack{\sigma\in\dop{\hh}:\tr{\rho\sigma}=0}}\lMnorm{\rho-\sigma},
\end{equation}
where we have used the fact $\tr{\rho\sigma}=0\Leftrightarrow\lpnorm{1}{\rho-\sigma}=2$. Note that $\mu_{\rho,\mathbb{M}}(1)$ is defined only for density operators $\rho$ that are not of full rank.

\section{Quantum Subspace Verification (QSV)}
Motivated by the scenario of fault-tolerant quantum computation where a logical state is encoded in a subspace, researchers have generalized quantum state verification into quantum subspace verification (QSV) \cite{FH17,CZZMZ24,ZXWZ25}. 
Here, we can formally define a QSV protocol as follows.
\begin{definition}
 For $\epsilon,\delta\in(0,1]$, an $(\epsilon,\delta)$-QSV protocol for a subspace $\vv\subseteq\hh$ consists of measurements performed on multiple copies $\rho^{\otimes m}$ of a tested state $\rho\in\dop{\hh}$ and a classical post-processing such that it outputs {\rm \textbf{Accept}} with probability at least $1-\delta$ if $\range{\rho}\subseteq\vv$ and outputs {\rm \textbf{Reject}} with probability at least $1-\delta$ if $\tr{\rho\Pi_\vv}\leq1-\epsilon$. The minimum number $m$ of copies (over all the QSV protocols) required to achieve the goal is called the sample complexity.
\end{definition}
Note that quantum state verification corresponds to the case of $\dim\vv=1$.
We only consider sequential measurements performed on each sample of $\rho$ (see Fig.~\ref{fig:QSVQDH} (a)), which is a reasonable assumption satisfied by most of the existing QSV protocols. We characterize the sample complexity for QSV by using a quantity defined as follows.
\begin{definition}
\label{def:gapQSV}
 For $\epsilon\in(0,1]$, a nontrivial proper subspace $(\{0\}\subsetneq)\vv(\subsetneq\hh)$, and a measurement class $\mathbb{M}$, we define an $\epsilon$-visibility $\gamma_{\vv,\mathbb{M}}(\epsilon)$ as follows.
 \begin{eqnarray}
    \gamma_{\vv,\mathbb{M}}(\epsilon)&:=&\frac{1}{\epsilon}\sup_{\Omega\in \mathbf{M}}\min_{\substack{\rho\in\dop{\hh}:\range{\rho}\subseteq\vv\\\sigma\in\dop{\hh}:\tr{\sigma\Pi_\vv}\leq1-\epsilon}}\tr{\Omega(\rho-\sigma)},
\end{eqnarray}
where $\mathbf{M}$ is the set of binary measurements associated with $\mathbb{M}$ (see Definition \ref{def:binaryM}).
\end{definition}
Note that $\Omega$ in the supremum is often called a strategy\footnote{Several studies in the literature call $\Omega$ a strategy under the assumption of $\tr{\rho\Omega}=1$ for any $\rho$ such that $\range{\rho}\subseteq\vv$. However, we do not assume such a condition.} \cite{PLM18} since it plays a central role in the design of many QSV protocols.
We can show that the sample complexity based on sequential measurements is bounded as $\Omega( (\gamma_{\vv,\mathbb{M}}(\epsilon)\epsilon)^{-1}\log\left(\frac{1}{\delta}\right))$ and $O( (\gamma_{\vv,\mathbb{M}}(\epsilon)\epsilon)^{-2}\log\left(\frac{1}{\delta}\right))$ when $\epsilon$ and $\delta$ approach $0$, as shown in Appendix \ref{appendix:samplecomplexity}.

A QSV protocol corresponding to a specific parameter $\epsilon(>0)$ can be constructed by using the optimized strategy $\Omega_\epsilon$ appearing in Definition \ref{def:gapQSV} for $\epsilon$, as shown in Proposition \ref{prop:samplecomplexity}; however, $\Omega_\epsilon$ may not be useful for constructing a QSV protocol for smaller $\epsilon'\in(0,\epsilon)$. In contrast, many QSV protocols use a strategy $\Omega$ that is applicable to any positive value of $\epsilon(>0)$. We refer to such an $\Omega$ as a universal strategy.
In the following, we define a fundamental quantity that determines the sample complexity required for verifying a subspace using universal strategies.
\begin{definition}
For a nontrivial proper subspace $(\{0\}\subsetneq)\vv(\subsetneq\hh)$ and a measurement class $\mathbb{M}$, we define the spectral gap $\hat{\gamma}_{\vv,\mathbb{M}}$ of the universal strategy as follows:
\begin{eqnarray}
   \hat{\gamma}_{\vv,\mathbb{M}}&:=&\sup_{\substack{\Omega\in \mathbf{M}\\\Omega\Pi_\vv\propto\Pi_\vv}}\min_{\substack{\rho\in\dop{\hh}:\range{\rho}\subseteq\vv\\\sigma\in\dop{\hh}:\range{\sigma}\subseteq\vv_\bot}}\tr{\Omega(\rho-\sigma)},
\end{eqnarray} 
where $\mathbf{M}$ is the set of binary measurements associated with $\mathbb{M}$ (see Definition \ref{def:binaryM}).
\end{definition}
Note that the condition $\Omega\Pi_\vv\propto\Pi_\vv$ is necessary for a strategy $\Omega$ to be universal.
This is because if $\Omega$ is a universal strategy, for any $\epsilon\in(0,1]$, there exists $s>0$ such that $\min_{\substack{\rho:\range{\rho}\subseteq\vv\\\sigma:\tr{\sigma\Pi_\vv}\leq1-\epsilon}}\tr{\Omega(\rho-\sigma)}>\epsilon s$. By using the following lemma, we obtain $\Omega\Pi_\vv\propto\Pi_\vv$.

\begin{lemma}
\label{lemma:deviation}
    Let $0\leq\Omega\leq\idop$ and $s,\epsilon\in(0,1)$. If 
    \begin{equation}
    \label{eq:epsvisible}
    \min_{\substack{\rho\in\dop{\hh}:\range{\rho}\subseteq\vv\\\sigma\in\dop{\hh}:\tr{\sigma\Pi_\vv}\leq1-\epsilon}}\tr{\Omega(\rho-\sigma)}>\epsilon s, 
\end{equation}
    \begin{equation}
       \min_{\substack{\rho\in\dop{\hh}:\range{\rho}\subseteq\vv\\\sigma\in\dop{\hh}:\range{\sigma}\subseteq\vv_\bot}}\tr{\Omega(\rho-\sigma)}>s,\ \lpnorm{\infty}{\Delta_1}<\frac{1}{2}\sqrt{\frac{\epsilon}{1-\epsilon}},\ \lpnorm{\infty}{\Delta_2}<\frac{\epsilon}{1-\epsilon},
    \end{equation}
    where $\lpnorm{\infty}{X}$ is the operator norm of $X$, $\Delta_1=(\idop-\Pi_\vv)\Omega\Pi_\vv$ and  $\Delta_2=\Pi_\vv\Omega\Pi_\vv-\frac{\tr{\Omega\Pi_\vv}}{\tr{\Pi_\vv}}\Pi_\vv$.
\end{lemma}
\begin{proof}
The first statement is obtained via the following calculation.
    \begin{eqnarray}
        &&(LHS\ of\ Eq.~\eqref{eq:epsvisible})\\
        &\leq&\min_{\substack{\rho\in\dop{\hh}:\range{\rho}\subseteq\vv\\\sigma\in\dop{\hh}:\range{\sigma}\subseteq \vv_\bot}}\tr{\Omega(\rho-((1-\epsilon)\rho+\epsilon\sigma))}\\
        &=&\min_{\substack{\rho\in\dop{\hh}:\range{\rho}\subseteq\vv\\\sigma\in\dop{\hh}:\range{\sigma}\subseteq \vv_\bot}}\epsilon\tr{\Omega(\rho-\sigma)}.
    \end{eqnarray}

The second statement is obtained via the following calculation.
    \begin{eqnarray}
        &&(LHS\ of\ Eq.~\eqref{eq:epsvisible})\\
        &\leq&\min_{\substack{\ket{\Psi}\in\vv\\\ket{\Psi_\bot}\in\vv_\bot}}\tr{\Omega\Psi}-\left(\sqrt{1-\epsilon}\bra{\Psi}+\sqrt{\epsilon}\bra{\Psi_\bot}\right)\Omega\left(\sqrt{1-\epsilon}\ket{\Psi}+\sqrt{\epsilon}\ket{\Psi_\bot}\right)\nonumber\\\\
        &=&\min_{\substack{\ket{\Psi}\in\vv\\\ket{\Psi_\bot}\in\vv_\bot}}\epsilon\tr{\Omega(\Psi-\Psi_\bot)}-2\sqrt{\epsilon(1-\epsilon)}|\bra{\Psi}\Omega\ket{\Psi_\bot}|\\
        &\leq&\epsilon-2\sqrt{\epsilon(1-\epsilon)}\lpnorm{\infty}{\Delta_1}.
    \end{eqnarray}

The third statement is obtained via the following calculation.

    \begin{eqnarray}
        &&(LHS\ of\ Eq.~\eqref{eq:epsvisible})\\
        &\leq&\min_{\substack{\ket{\Psi},\ket{\Phi}\in\vv\\\ket{\Psi_\bot}\in\vv_\bot}}\tr{\Omega\Phi}-\left(\sqrt{1-\epsilon}\bra{\Psi}+\sqrt{\epsilon}\bra{\Psi_\bot}\right)\Omega\left(\sqrt{1-\epsilon}\ket{\Psi}+\sqrt{\epsilon}\ket{\Psi_\bot}\right)\nonumber\\\\
        &\leq&\min_{\substack{\ket{\Phi},\ket{\Psi}\in\vv\\\ket{\Psi_\bot}\in\vv_\bot}}(1-\epsilon)\tr{\Omega(\Phi-\Psi)}+\epsilon\tr{\Omega(\Phi-\Psi_\bot)}\\
        &\leq&\epsilon-(1-\epsilon)\lpnorm{\infty}{\Delta_2}.
    \end{eqnarray}
    
\end{proof}

Note also that the sample complexity based on this universal strategy is upper bounded by $O( (\hat{\gamma}_{\vv,\mathbb{M}}\epsilon)^{-2}\log\left(\frac{1}{\delta}\right))$ when $\epsilon$ and $\delta$ approach $0$. This can be easily proven by following the argument in the proof of Proposition \ref{prop:samplecomplexity} in Appendix \ref{appendix:samplecomplexity} and utilizing the following Proposition.
\begin{proposition}
\label{prop:visibility_gap_bound}
 For a nontrivial proper subspace $(\{0\}\subsetneq)\vv(\subsetneq\hh)$, a real number $\epsilon\in(0,1]$, and a POVM element $\Omega$ such that $\Omega\Pi_\vv\propto\Pi_\vv$, it holds that
 \begin{eqnarray}
 \label{eq:strategy_from_universal}
   \min_{\substack{\rho\in\dop{\hh}:\range{\rho}\subseteq\vv\\\sigma\in\dop{\hh}:\range{\sigma}\subseteq\vv_\bot}}\tr{\Omega(\rho-\sigma)}\geq s
   \Rightarrow
     \min_{\substack{\rho\in\dop{\hh}:\range{\rho}\subseteq\vv\\\sigma\in\dop{\hh}:\tr{\sigma\Pi_\vv}\leq1-\epsilon}}\tr{\Omega(\rho-\sigma)}\geq \epsilon s. 
\end{eqnarray} 
Moreover, this implies $\hat{\gamma}_{\vv,\mathbb{M}}\leq \gamma_{\vv,\mathbb{M}}(\epsilon)$ for any $\epsilon\in(0,1]$.
\end{proposition}
\begin{proof}
 The above statement $\hat{\gamma}_{\vv,\mathbb{M}}\leq \gamma_{\vv,\mathbb{M}}(\epsilon)$ is simply obtained by using Eq.~\eqref{eq:strategy_from_universal} in the definition of $\hat{\gamma}_{\vv,\mathbb{M}}$ and $\gamma_{\vv,\mathbb{M}}(\epsilon)$. Eq.~\eqref{eq:strategy_from_universal} can be obtained as follows.
Since $\Omega\Pi_\vv\propto\Pi_\vv$, $\min_{\substack{\rho\in\dop{\hh}:\range{\rho}\subseteq\vv\\\sigma\in\dop{\hh}:\range{\sigma}\subseteq\vv_\bot}}\tr{\Omega(\rho-\sigma)}\geq s$ implies that $\min_{\substack{\rho:\range{\rho}\subseteq\vv\\\sigma:\range{\sigma}\subseteq\vv_\bot}}\tr{\Omega(\rho-\sigma)}=\lambda_d(\Omega)-\lambda_{d+1}(\Omega)\geq s$ and $\lambda_1(\Omega)=\lambda_2(\Omega)=\cdots=\lambda_d(\Omega)$, where $\lambda_i$ is the $i$-th largest eigenvalue and $\dim\vv=d$. In this case, we obtain
    \begin{eqnarray}
        \min_{\substack{\rho:\range{\rho}\subseteq\vv\\\sigma:\tr{\sigma\Pi_\vv}\leq1-\epsilon}}\tr{\Omega(\rho-\sigma)}
        =\epsilon(\lambda_d(\Omega)-\lambda_{d+1}(\Omega))
        \geq\epsilon s.
    \end{eqnarray}
\end{proof}

\section{Properties of fundamental quantities}
In this section, we show that the fundamental quantities described above have the properties listed in Table~\ref{tab:property}.
\begin{table}[h]
\centering
\begin{tabular}{|c|c|c|c|}
\hline
quantity & monotonicity & continuity & limit value $(\epsilon\rightarrow0)$ \\
\hline
$\epsilon$-distinguishability ratio $\mu_{\rho,\mathbb{M}}(\epsilon)$ & 
 YES [Proposition \ref{prop:monotone_distratio}] & N/A & distinguishability ratio $\hat{\mu}_{\rho,\mathbb{M}}$ [Proposition \ref{prop:limit_distratio}] \\
$\epsilon$-visibility $\gamma_{\vv,\mathbb{M}}(\epsilon)$ & YES [Proposition \ref{prop:increasing_visibility}] & YES [Proposition \ref{prop:continuity_visibility}] & spectral gap $\hat{\gamma}_{\vv,\mathbb{M}}$ [Proposition \ref{prop:QSV}] \\
\hline
\end{tabular}
\caption{Properties of the fundamental quantities for QSV and QDH. Monotonicity means that the function is monotonically non-decreasing. Since we do not use the continuity of $\gamma_{\vv,\mathbb{M}}(\epsilon)$ in the subsequent sections, we will leave its proof to Appendix \ref{appendix:continuity}.
Moreover, we will not concern ourselves with the continuity of $\mu_{\rho,\mathbb{M}}(\epsilon)$. Indeed, it may not be continuous in the general case.}
\label{tab:property}
\end{table}

\begin{proposition}
\label{prop:monotone_distratio}
 For a quantum state $\rho\in\dop{\hh}$ and a measurement class $\mathbb{M}$, the $\epsilon$-distinguishability ratio $\mu_{\rho,\mathbb{M}}(\epsilon)$ is a monotonically non-decreasing function in $\epsilon\in(0,\epsilon_\rho]$. 
\end{proposition}
\begin{proof}
 For $\epsilon,\epsilon'\in(0,\epsilon_\rho]$ such that $\epsilon<\epsilon'$, let $\mu_{\rho,\mathbb{M}}(\epsilon')=\frac{1}{2\epsilon'}\lMnorm{\rho-\hat{\sigma}_{\epsilon'}}$, where $\hat{\sigma}_{\epsilon'}$ satisfies $\lpnorm{1}{\rho-\hat{\sigma}_{\epsilon'}}\geq2\epsilon'$. By letting $\sigma_{\epsilon}=(1-p)\hat{\sigma}_{\epsilon'}+p\rho$, where $p=1-\frac{\epsilon}{\epsilon'}$, we can verify that
 \begin{eqnarray}
 \lpnorm{1}{\rho-\sigma_{\epsilon}}=(1-p) \lpnorm{1}{\rho-\hat{\sigma}_{\epsilon'}}\geq2(1-p)\epsilon'=2\epsilon
\end{eqnarray}
and
\begin{eqnarray}
 \mu_{\rho,\mathbb{M}}(\epsilon)\leq\frac{1}{2\epsilon}\lMnorm{\rho-\sigma_{\epsilon}}=\frac{1-p}{2\epsilon}\lMnorm{\rho-\hat{\sigma}_{\epsilon'}}= \mu_{\rho,\mathbb{M}}(\epsilon').
\end{eqnarray}
This completes the proof.
\end{proof}

\begin{proposition}
\label{prop:limit_distratio}
 For a quantum state $\rho\in\dop{\hh}$ and a measurement class $\mathbb{M}$, the $\epsilon$-distinguishability ratio $\mu_{\rho,\mathbb{M}}(\epsilon)$ satisfies
\begin{equation}
 \lim_{\epsilon\rightarrow0}\mu_{\rho,\mathbb{M}}(\epsilon)=\hat{\mu}_{\rho,\mathbb{M}},
\end{equation}
where $\hat{\mu}_{\rho,\mathbb{M}}$ is the distinguishability ratio of Definition \ref{def:distratio}
\end{proposition}
\begin{proof}
Since $\mu_{\rho,\mathbb{M}}(\epsilon)$ is non-decreasing and is lower bounded by $0$, its limit $\lim_{\epsilon\rightarrow0}\mu_{\rho,\mathbb{M}}(\epsilon)$ exists.
 By definition,
 \begin{equation}
 \mu_{\rho,\mathbb{M}}(\epsilon)\geq\frac{1}{2\epsilon}\min_{\substack{\sigma\in\dop{\hh}:\lpnorm{1}{\rho-\sigma}\geq2\epsilon}}\hat{\mu}_{\rho,\mathbb{M}}\lpnorm{1}{\rho-\sigma}=\hat{\mu}_{\rho,\mathbb{M}}
\end{equation}
 for any $\epsilon$. This implies $(LHS)\geq (RHS)$.
 On the other hand, for any real number $c>\hat{\mu}_{\rho,\mathbb{M}}$, there exists a state $\hat{\sigma}$ such that $\lMnorm{\rho-\hat{\sigma}}< c\lpnorm{1}{\rho-\hat{\sigma}}$.
 By letting $\epsilon=\frac{1}{2}\lpnorm{1}{\rho-\hat{\sigma}}$, we can verify that
 \begin{equation}
 \mu_{\rho,\mathbb{M}}(\epsilon)\leq\frac{1}{2\epsilon}\lMnorm{\rho-\hat{\sigma}}<\frac{c}{2\epsilon}\lpnorm{1}{\rho-\hat{\sigma}}=c.
\end{equation}
This implies $(LHS)\leq (RHS)$.
\end{proof}
This proposition justifies the notion of the $\epsilon$-distinguishability ratio $\mu_{\rho,\mathbb{M}}(\epsilon)$ as a generalization of the distinguishability ratio $\hat{\mu}_{\rho,\mathbb{M}}$.

From Proposition \ref{prop:monotone_distratio} and Proposition \ref{prop:limit_distratio}, we find that $\hat{\mu}_{\rho,\mathbb{M}}\leq \mu_{\rho,\mathbb{M}}(1)$.
One would wonder if there is a lower bound on $\hat{\mu}_{\rho,\mathbb{M}}$ of the form $c\mu_{\rho,\mathbb{M}}(1)\leq\hat{\mu}_{\rho,\mathbb{M}}$ with a positive constant $c>0$ independent of $\rho$ and $\mathbb{M}$.
In the following, we give an example showing that such a lower bound does not exist.

\noindent\textbf{Example:} Let $\ket{\Phi}\in\vee_2\cd$ be a state in the symmetric subspace $\vee_2\cd:=\{\ket{\tilde{\Phi}}\in\cd\otimes\cd:U_{SWAP}\ket{\tilde{\Phi}}=\ket{\tilde{\Phi}}\}$, where $U_{SWAP}$ represents the swap operator. By letting $\rho=\frac{1}{d^2-1}(\idop-\Phi)$, we find that
\begin{equation}
 \mu_{\rho,\mathbb{LO}}(1)=\frac{1}{2}\lpnorm{\mathbb{LO}}{\rho-\Phi}\geq\frac{\hat{\mu}_{\Phi,\mathbb{LO}}}{2}\lpnorm{1}{\rho-\Phi}\geq\frac{1}{2\sqrt{153}}
\end{equation}
where we have used Eq.~\eqref{eq:2LOratio} and the orthogonality between $\rho$ and $\Phi$ to derive the last inequality.

On the other hand, by letting $\sigma=\frac{1}{d^2-1}(\frac{2d}{d+1}\Pi_{\vee_2\cd}-\Phi)$, we obtain
\begin{eqnarray}
 \hat{\mu}_{\rho,\mathbb{LO}}&\leq&\frac{\lpnorm{\mathbb{LO}}{\rho-\sigma}}{\lpnorm{1}{\rho-\sigma}}=\frac{1}{2}\lpnorm{\mathbb{LO}}{\alpha\Pi_{\wedge_2\cd}-\beta\Pi_{\vee_2\cd}},
\end{eqnarray}
where we have used $\rho-\sigma=\frac{d}{2(d+1)}\left(\alpha\Pi_{\wedge_2\cd}-\beta\Pi_{\vee_2\cd}\right)$, $\alpha=(\dim\wedge_2\cd)^{-1}=\frac{2}{d(d-1)}$ and $\beta=(\dim\vee_2\cd)^{-1}=\frac{2}{d(d+1)}$. Note that $\wedge_2\cd:=\{\ket{\tilde{\Phi}}\in\cd\otimes\cd:U_{SWAP}\ket{\tilde{\Phi}}=-\ket{\tilde{\Phi}}\}$ is an anti-symmetric subspace.
 We can proceed with the calculation as follows.
\begin{eqnarray}
\label{eq:SymPPTbound}
 \lpnorm{\mathbb{LO}}{\alpha\Pi_{\wedge_2\cd}-\beta\Pi_{\vee_2\cd}}&\leq& \lpnorm{\mathbb{PPT}}{\alpha\Pi_{\wedge_2\cd}-\beta\Pi_{\vee_2\cd}}\\
 &=& \lpnorm{\mathbb{PPT}}{\alpha\Pi_{\wedge_2\cd}^{T_1}-\beta\Pi_{\vee_2\cd}^{T_1}}\\
 &\leq& \lpnorm{1}{\alpha\Pi_{\wedge_2\cd}^{T_1}-\beta\Pi_{\vee_2\cd}^{T_1}}\\
 &=&\frac{1}{2}\lpnorm{1}{\alpha(\idop-U_{SWAP})^{T_1}-\beta(\idop+U_{SWAP})^{T_1}}\\
 &=&\frac{2}{d^2-1}\lpnorm{1}{\frac{1}{d}\idop-d\Phi^+}=\frac{4}{d},
\end{eqnarray}
where $\mathbb{PPT}$ is a measurement class realizable as bipartite PPT measurements, $T_1$ represents the partial transposition on the first qudit, and $\Phi^+$ is a maximally entangled state.
Thus, we obtain $\hat{\mu}_{\rho,\mathbb{LO}}\leq\frac{2}{d}$, which cannot be lower bounded by $c\mu_{\rho,\mathbb{LO}}(1)$ with a positive constant $c$.

We can show the following property about $\epsilon$-visibility.
\begin{proposition}
\label{prop:increasing_visibility}
  For a nontrivial proper subspace $(\{0\}\subsetneq)\vv(\subsetneq\hh)$ and a measurement class $\mathbb{M}$, an $\epsilon$-visibility $\gamma_{\vv,\mathbb{M}}(\epsilon)$ is a monotonically non-decreasing function in $\epsilon\in(0,1]$.
\end{proposition}
\begin{proof}
 Suppose $0<\epsilon<\epsilon'\leq1$. By definition, for any $\delta>0$, there exists $\Omega\in\mathbf{M}$ such that
\begin{equation}
 \min_{\substack{\rho\in\dop{\hh}:\range{\rho}\subseteq\vv\\\sigma\in\dop{\hh}:\tr{\sigma\Pi_\vv}\leq1-\epsilon}}\tr{\Omega(\rho-\sigma)}>\epsilon \gamma_{\vv,\mathbb{M}}(\epsilon)-\delta.
\end{equation}
For any states $\rho\in\dop{\hh}$ and $\sigma\in\dop{\hh}$ such that $\range{\rho}\subseteq\vv$ and $\tr{\sigma\Pi_\vv}\leq1-\epsilon'$, we can verify that $\hat{\sigma}=\frac{\epsilon'-\epsilon}{\epsilon'}\rho+\frac{\epsilon}{\epsilon'}\sigma$ satisfies 
\begin{equation}
\tr{\hat{\sigma}\Pi_\vv}=\frac{\epsilon'-\epsilon}{\epsilon'}+\frac{\epsilon}{\epsilon'}\tr{\sigma\Pi_\vv}\leq1-\epsilon.
\end{equation}
This implies that 
\begin{eqnarray}
 \epsilon' \gamma_{\vv,\mathbb{M}}(\epsilon')&\geq&\min_{\substack{\rho\in\dop{\hh}:\range{\rho}\subseteq\vv\\\sigma\in\dop{\hh}:\tr{\sigma\Pi_\vv}\leq1-\epsilon'}}\tr{\Omega(\rho-\sigma)}\\
 &=&\min_{\substack{\rho\in\dop{\hh}:\range{\rho}\subseteq\vv\\\sigma\in\dop{\hh}:\tr{\sigma\Pi_\vv}\leq1-\epsilon'}}\frac{\epsilon'}{\epsilon}\tr{\Omega(\rho-\hat{\sigma})}\\
& >&\frac{\epsilon'}{\epsilon}(\epsilon \gamma_{\vv,\mathbb{M}}(\epsilon)-\delta)=\epsilon'\left(\gamma_{\vv,\mathbb{M}}(\epsilon)-\frac{\delta}{\epsilon}\right).
\end{eqnarray}
This completes the proof.
\end{proof}

While a universal strategy seems to be much more restrictive than one that would depend on $\epsilon$,
we can show that their performance indicators, $\hat{\gamma}_{\vv,\mathbb{M}}$ and $\gamma_{\vv,\mathbb{M}}(\epsilon)$, are smoothly connected in the following way.
\begin{proposition}
\label{prop:QSV}
For a nontrivial proper subspace $(\{0\}\subsetneq)\vv(\subsetneq\hh)$ and a measurement class $\mathbb{M}$ that is informationally complete, the $\epsilon$-visibility $\gamma_{\vv,\mathbb{M}}(\epsilon)$ and the spectral gap $\hat{\gamma}_{\vv,\mathbb{M}}$ satisfy
\begin{equation}
\label{eq:QSVlimit}
  \hat{\gamma}_{\vv,\mathbb{M}}=\lim_{\epsilon\rightarrow0}\gamma_{\vv,\mathbb{M}}(\epsilon). 
\end{equation}
\end{proposition}

\begin{proof}
 Since $\gamma_{\vv,\mathbb{M}}(\epsilon)$ is non-negative and is monotonically non-decreasing, its limit $s:=\lim_{\epsilon\rightarrow0}\gamma_{\vv,\mathbb{M}}(\epsilon)$ exists. Since Proposition \ref{prop:visibility_gap_bound} implies that $0\leq\hat{\gamma}_{\vv,\mathbb{M}}\leq s\leq \gamma_{\vv,\mathbb{M}}(1)\leq1$, we obtain $s\in[0,1]$. If $s=0$, we can verify that $\hat{\gamma}_{\vv,\mathbb{M}}=0$, which implies Eq.~\eqref{eq:QSVlimit}.
On the other hand, if $s>0$, we find that for any $s'\in(0,s)$ and $\epsilon\in(0,1]$, $\gamma_{\vv,\mathbb{M}}(\epsilon)>s'$. This implies that for any $\epsilon\in(0,1]$, there exists a binary measurement $\Omega(\epsilon)\in\mathbf{M}$ such that 
 \begin{equation}
 \label{eq:strategycondition}
 	\min_{\substack{\rho\in\dop{\hh}:\range{\rho}\subseteq\vv\\\sigma\in\dop{\hh}:\tr{\sigma\Pi_\vv}\leq1-\epsilon}}\tr{\Omega(\epsilon)(\rho-\sigma)}>\epsilon s'.
\end{equation}
 
Since Lemma \ref{lemma:deviation} guarantees that $\Omega(\epsilon)$ is nearly a universal strategy for small $\epsilon$, we can construct a universal strategy $\hat{\Omega}$ from a convex combination $\hat{\Omega}=(1-p)\Omega(\epsilon)+p\Omega'$ where the parameters $p,\epsilon\in(0,1)$ and $\Omega'\in\mathbf{M}$ are chosen as follows.
    By letting $\Delta_1:=(\idop-\Pi_\vv)\Omega(\epsilon)\Pi_\vv$ and $\Delta_2:=\Pi_\vv\Omega(\epsilon)\Pi_\vv-\frac{\tr{\Omega(\epsilon)\Pi_\vv}}{\tr{\Pi_\vv}}\Pi_\vv$, we obtain
    \begin{eqnarray}
        \hat{\Omega}\Pi_\vv&=&(1-p)\Omega(\epsilon)\Pi_\vv+p\Omega'\Pi_\vv\\
        &=&(1-p)\Pi_\vv\Omega(\epsilon)\Pi_\vv+(1-p)\Delta_1+p\Omega'\Pi_\vv\\
        &=&(1-p)\frac{\tr{\Omega(\epsilon)\Pi_\vv}}{\tr{\Pi_\vv}}\Pi_\vv+(1-p)(\Delta_1+\Delta_2)+p\Omega'\Pi_\vv.
    \end{eqnarray}
    By setting $\Omega'=\frac{1}{2}\idop-\frac{1-p}{p}(\Delta_1+\Delta_1^\dag+\Delta_2)$, we find that $\hat{\Omega}\Pi_\vv\propto\Pi_\vv$.
    By setting $p=\epsilon^{\frac{1}{3}}$ and applying Lemma \ref{lemma:deviation} to Eq.~\eqref{eq:strategycondition},  we obtain $\frac{1-p}{p}\lpnorm{\infty}{\Delta_1+\Delta_1^\dag+\Delta_2}\leq\epsilon^{-\frac{1}{3}}(2\lpnorm{\infty}{\Delta_1}+\lpnorm{\infty}{\Delta_2})\leq2\sqrt{2}\epsilon^{\frac{1}{6}}$ if $\epsilon\in\left(0,\frac{1}{2}\right]$.
    Lemma \ref{lemma:ball} in Appendix \ref{appendix:eball} implies that $\Omega'\in\mathbf{M}$ for sufficiently small $\epsilon$.

    Suppose that $\epsilon$ is sufficiently small such that $\Omega'\in\mathbf{M}$. Since $\mathbf{M}$ is convex, we have that $\hat{\Omega}\in\mathbf{M}$.
    This implies that
\begin{eqnarray}
 	\hat{\gamma}_{\vv,\mathbb{M}}&\geq&\min_{\substack{\rho\in\dop{\hh}:\range{\rho}\subseteq\vv\\\sigma\in\dop{\hh}:\range{\sigma}\subseteq\vv_\bot}}\tr{\hat{\Omega}(\rho-\sigma)}\\
	&\geq&\min_{\substack{\rho\in\dop{\hh}:\range{\rho}\subseteq\vv\\\sigma\in\dop{\hh}:\range{\sigma}\subseteq\vv_\bot}}(1-p)\tr{\Omega(\epsilon)(\rho-\sigma)}\\
	&&+\min_{\substack{\rho\in\dop{\hh}:\range{\rho}\subseteq\vv\\\sigma\in\dop{\hh}:\range{\sigma}\subseteq\vv_\bot}}p\tr{\Omega'(\rho-\sigma)}\\
	&\geq&\min_{\substack{\rho\in\dop{\hh}:\range{\rho}\subseteq\vv\\\sigma\in\dop{\hh}:\range{\sigma}\subseteq\vv_\bot}}(1-p)\tr{\Omega(\epsilon)(\rho-\sigma)}-p.
\end{eqnarray}
    By applying Lemma \ref{lemma:deviation} to Eq.~\eqref{eq:strategycondition}, we obtain $\hat{\gamma}_{\vv,\mathbb{M}}\geq (1-p)s'-p=(1-\epsilon^{\frac{1}{3}})s'-\epsilon^{\frac{1}{3}}$.
    Since this inequality holds for any $s'<\lim_{\epsilon\rightarrow0}\gamma_{\vv,\mathbb{M}}(\epsilon)$ and sufficiently small $\epsilon$,
    $\hat{\gamma}_{\vv,\mathbb{M}}\geq\lim_{\epsilon\rightarrow0}\gamma_{\vv,\mathbb{M}}(\epsilon)$. Combining the above with Proposition \ref{prop:visibility_gap_bound} completes the proof.
    \end{proof}

\section{Duality between QSV and QDH}
In this section, we prove the duality between the hardness of QSV under a measurement class $\mathbb{M}$ and the security of QDH against malicious data servers capable of performing a measurement in $\mathbb{M}$, as mentioned in the introduction.
Sion's minimax theorem and several properties of the fundamental quantities of QSV and QDH described in the previous sections play key roles in proving this duality.

First, we provide a dual expression for $\epsilon$-visibility, which helps to establish its connection to the $\epsilon$-distinguishability ratio.
\begin{proposition}
\label{prop:visibilityminimax}
 For $\epsilon\in(0,1]$, a nontrivial proper subspace $(\{0\}\subsetneq)\vv(\subsetneq\hh)$, and a measurement class $\mathbb{M}$, $\epsilon$-visibility $\gamma_{\vv,\mathbb{M}}(\epsilon)$ satisfies
\begin{equation}
\label{eq:minimaxform}
 \gamma_{\vv,\mathbb{M}}(\epsilon)=\min_{\substack{\rho\in\dop{\hh}:\range{\rho}\subseteq\vv\\\sigma\in\dop{\hh}:\tr{\sigma\Pi_\vv}\leq1-\epsilon}}\frac{1}{2\epsilon}\lMnorm{\rho-\sigma}.
\end{equation}
\end{proposition}
\begin{proof}
Eq.~\eqref{eq:minimaxform} can be derived from the following calculation.
\begin{eqnarray}
    \epsilon \gamma_{\vv,\mathbf{M}}(\epsilon)&=&\sup_{\Omega\in \mathbf{M}}\min_{\substack{\rho\in\dop{\hh}:\range{\rho}\subseteq\vv\\\sigma\in\dop{\hh}:\tr{\sigma\Pi_\vv}\leq1-\epsilon}}\tr{\Omega(\rho-\sigma)}\\
    &=&\sup_{\Omega\in \mathbf{M}}\min_{\eta\in\mathbf{S}_\epsilon}\tr{\Omega(\ptr{2}{\eta}-\ptr{1}{\eta})}\\
    &=&\min_{\eta\in\mathbf{S}_\epsilon}\sup_{\Omega\in \mathbf{M}}\tr{\Omega(\ptr{2}{\eta}-\ptr{1}{\eta})}\\
    &=&\min_{\substack{\rho\in\dop{\hh}:\range{\rho}\subseteq\vv\\\sigma\in\dop{\hh}:\tr{\sigma\Pi_\vv}\leq1-\epsilon}}\frac{1}{2}\lMnorm{\rho-\sigma},
\end{eqnarray}
where  $\textbf{S}_\epsilon:={\rm conv}(\{\rho\otimes\sigma:\range{\rho}\subseteq\vv,\tr{\sigma\Pi_\vv}\leq1-\epsilon\})$ is a compact convex subset of $\dop{\hh_1\otimes\hh_2}$, $\hh_1\simeq\hh_2\simeq\hh$ and we have used Sion's minimax theorem in the third equality since $f(\Omega,\eta)=\tr{\Omega(\ptr{2}{\eta}-\ptr{1}{\eta})}$ is bilinear. 
\end{proof}

Second, we show a relationship between the $\epsilon$-visibility and the $\epsilon$-distinguishability ratio.
\begin{theorem}
\label{thm:comparison}
  For a nontrivial proper subspace $(\{0\}\subsetneq)\vv(\subsetneq\hh)$ and a measurement class $\mathbb{M}$, $\epsilon$-visibility $\gamma_{\vv,\mathbb{M}}(\epsilon)$ and $\epsilon$-distinguishability ratio $\mu_{\rho,\mathbb{M}}(\epsilon)$ satisfy that
  \begin{equation}
  \label{eq:rs}
 	\mu_{\vv,\mathbb{M}}(\epsilon):=\min_{\rho\in\dop{\hh}:\range{\rho}\subseteq\vv}\mu_{\rho,\mathbb{M}}(\epsilon)\leq \gamma_{\vv,\mathbb{M}}(\epsilon)
\end{equation}
for any $\epsilon\in(0,1]$.
Moreover, their limiting quantities, i.e., the distinguishability ratio $\hat{\mu}_{\rho,\mathbb{M}}$ and the spectral gap $\hat{\gamma}_{\vv,\mathbb{M}}$, satisfy
\begin{equation}
  \label{eq:rslimit}
 	\hat{\mu}_{\vv,\mathbb{M}}:=\inf_{\rho\in\dop{\hh}:\range{\rho}\subseteq\vv}\hat{\mu}_{\rho,\mathbb{M}}\leq \hat{\gamma}_{\vv,\mathbb{M}}
\end{equation}
if $\mathbb{M}$ is informationally complete.
\end{theorem}

\begin{proof}
Since $\lpnorm{1}{\rho-\sigma}\geq2\tr{\Pi_\vv(\rho-\sigma)}\geq2\epsilon$ if $\range{\rho}\subseteq\vv$ and $\tr{\sigma\Pi_\vv}\leq1-\epsilon$, we obtain
 \begin{eqnarray}
  \gamma_{\vv,\mathbb{M}}(\epsilon)&=&\min_{\substack{\rho\in\dop{\hh}:\range{\rho}\subseteq\vv\\\sigma\in\dop{\hh}:\tr{\sigma\Pi_\vv}\leq1-\epsilon}}\frac{1}{2\epsilon}\lMnorm{\rho-\sigma}\geq
  \min_{\substack{\rho\in\dop{\hh}:\range{\rho}\subseteq\vv\\\sigma\in\dop{\hh}:\lpnorm{1}{\rho-\sigma}\geq2\epsilon}}\frac{1}{2\epsilon}\lMnorm{\rho-\sigma}
 =\min_{\rho\in\dop{\hh}:\range{\rho}\subseteq\vv}\mu_{\rho,\mathbb{M}}(\epsilon),
\end{eqnarray}
where we have used Eq.~\eqref{eq:minimaxform} to derive the first equality. Note that we can take the minimization of $\frac{1}{2\epsilon}\lMnorm{\rho-\sigma}$ over $\rho,\sigma$ satisfying $\range{\rho}\subseteq\vv$ and $\lpnorm{1}{\rho-\sigma}\geq2\epsilon$ since the semi-norm is a continuous function and the region of $\rho,\sigma$ is compact.

The second statement can be obtained via the following calculation.
\begin{eqnarray}
 \hat{\gamma}_{\vv,\mathbb{M}}&=&\inf_{\epsilon\in(0,1]}\gamma_{\vv,\mathbb{M}}(\epsilon)\\
 &\geq&\inf_{\epsilon\in(0,1]}\min_{\rho\in\dop{\hh}:\range{\rho}\subseteq\vv}\mu_{\rho,\mathbb{M}}(\epsilon)\\
 &\geq&\inf_{\epsilon\in(0,1]}\inf_{\rho\in\dop{\hh}:\range{\rho}\subseteq\vv}\hat{\mu}_{\rho,\mathbb{M}}=\inf_{\rho\in\dop{\hh}:\range{\rho}\subseteq\vv}\hat{\mu}_{\rho,\mathbb{M}},
\end{eqnarray}
where we have used Proposition \ref{prop:limit_distratio}, Proposition \ref{prop:QSV}, and the monotonically non-decreasing property of $\gamma_{\vv,\mathbb{M}}(\epsilon)$ and $\mu_{\rho,\mathbb{M}}(\epsilon)$ to derive the first equation and the last inequality.
\end{proof}

\begin{figure}[ht]
    \centering
    \includegraphics[width=18cm]{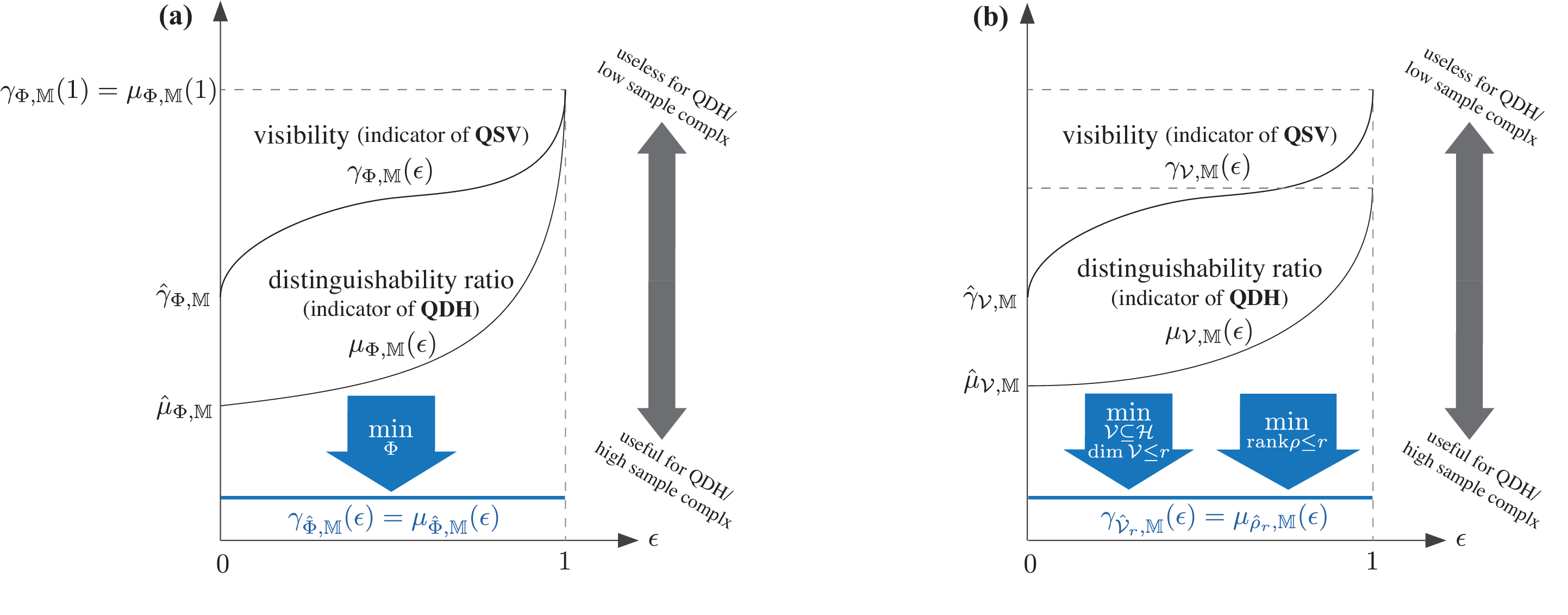}
    \caption{Relationship between fundamental quantities in QSV and QDH under a measurement class $\mathbb{M}$ that is informationally complete. 
    $\epsilon$-visibility $\gamma_{\vv,\mathbb{M}}(\epsilon)$ characterizes the sample complexity for verifying a subspace $\vv$ and $\epsilon$-distinguishability ratio $\mu_{\rho,\mathbb{M}}(\epsilon)$ characterizes the security of QDH using a quantum state $\rho$ and its best counterpart. As their limit values, we obtain the spectral gap $\hat{\gamma}_{\vv,\mathbb{M}}$ and the distinguishability ratio $\hat{\mu}_{\rho,\mathbb{M}}$. $\mu_{\vv,\mathbb{M}}(\epsilon)$ and $\hat{\mu}_{\vv,\mathbb{M}}$ are defined as the infimum of $\mu_{\rho,\mathbb{M}}(\epsilon)$ and $\hat{\mu}_{\rho,\mathbb{M}}$ over all the states $\rho$ such that $\range{\rho}\subseteq\vv$.
All the quantities coincide when we consider the most difficult subspaces for QSV and the most secure states for QDH.}
    \label{fig:diagram}
\end{figure}

In general, $\mu_{\vv,\mathbb{M}}(\epsilon)$ and $\gamma_{\vv,\mathbb{M}}(\epsilon)$ do not coincide for any $\epsilon\in(0,1]$ (see Fig.~\ref{fig:diagram}).
However, we can verify that $\mu_{\vv,\mathbb{M}}(1)=\gamma_{\vv,\mathbb{M}}(1)$ when $\dim\vv=1$ by comparing Eq.~\eqref{eq:distratio1} and Eq.~\eqref{eq:minimaxform}. 
Proving the statement $\hat{\mu}_{\vv,\mathbb{M}}\leq \hat{\gamma}_{\vv,\mathbb{M}}$ in the Theorem is nontrivial since only universal strategies, i.e., $\Omega\in\mathbb{M}$ satisfying $\Omega\Pi_\vv\propto\Pi_\vv$, are allowed in the definition of $\hat{\gamma}_{\vv,\mathbb{M}}$. In contrast, arbitrary POVMs $\{\Omega,\idop-\Omega\}$ with $\Omega\in\mathbf{M}$, which are not necessarily universal strategies, are allowed in the definition of $\hat{\mu}_{\vv,\mathbb{M}}$, which potentially increases the distinguishability ratio.

We can verify that Theorem \ref{thm:comparison} implies the first statement on the duality between QSV and QDH mentioned in the introduction as follows:
If for a fixed $\epsilon$, there does not exist an $(\epsilon,\delta)$-data hiding pair of states including any $\rho$ lying in $\vv$ against a measurement class $\mathbb{M}$ for any $\delta<\mu_0\epsilon$ (with a large number $\mu_0$), we find that $\mu_{\vv,\mathbb{M}}(\epsilon)\geq\mu_0$. Thus, Eq.~\eqref{eq:rs} implies $\gamma_{\vv,\mathbb{M}}(\epsilon)\geq\mu_0$. This guarantees the existence of an $(\epsilon,\delta)$-QSV protocol that can verify a subspace $\vv$ within $O( (\mu_0\epsilon)^{-2}\log\left(\frac{1}{\delta}\right))$ samples.

Third, we show how to orthogonalize two distinct states $\rho$ and $\sigma$ without changing the ratio $\frac{\lMnorm{\rho-\sigma}}{\lpnorm{1}{\rho-\sigma}}$.
\begin{theorem}
 \label{thm:constructionQDHfromQSV}
	For a set $\{\rho,\sigma\}\subseteq\dop{\hh}$ of distinct states  and a measurement class $\mathbb{M}$, there exists a set $\{\hat{\rho},\hat{\sigma}\}\subseteq\dop{\hh}$ of orthogonal states such that $\rank{\hat{\rho}}\leq\min\{\rank{\rho},\frac{\dim\hh}{2}\}$ and
\begin{equation}
\label{eq:QDHfromLDR}
\frac{1}{2}\lMnorm{\hat{\rho}-\hat{\sigma}}= \frac{\lMnorm{\rho-\sigma}}{\lpnorm{1}{\rho-\sigma}}.
\end{equation}
 Moreover,  for any natural number $r\leq\dim\hh$, it holds that
    	\begin{equation}
	\label{eq:distratio_QDH}
 		\min_{\substack{\rho\in\dop{\hh}\\\rank{\rho}\leq \min\{r,\frac{\dim\hh}{2}\}}}\mu_{\rho,\mathbb{M}}(1)=\min_{\substack{\rho\in\dop{\hh}\\\rank{\rho}\leq r}}\mu_{\rho,\mathbb{M}}(1)=\min_{\substack{\rho\in\dop{\hh}\\\rank{\rho}\leq r}}\hat{\mu}_{\rho,\mathbb{M}}.
	\end{equation}
\end{theorem}
\begin{proof}
    By decomposing $\rho-\sigma$ into $\lambda\left(\hat{\rho}-\hat{\sigma}\right)$, where $\lambda>0$ and $\hat{\rho}$ and $\hat{\sigma}$ are orthogonal mixed states, we obtain $\rank{\hat{\rho}}\leq \rank{\rho}$ and $\lMnorm{\rho-\sigma}=\lambda\lMnorm{\hat{\rho}-\hat{\sigma}}$ for any measurement class $\mathbb{M}$. This implies 
    \begin{eqnarray}
\frac{\lMnorm{\rho-\sigma}}{\lpnorm{1}{\rho-\sigma}}=\frac{\lambda\lMnorm{\hat{\rho}-\hat{\sigma}}}{\lambda\lpnorm{1}{\hat{\rho}-\hat{\sigma}}}=\frac{1}{2}\lMnorm{\hat{\rho}-\hat{\sigma}},
\end{eqnarray}
which in turn implies Eq.~\eqref{eq:QDHfromLDR} if $\rank{\hat{\rho}}\leq\frac{\dim\hh}{2}$. Otherwise, we can verify Eq.~\eqref{eq:QDHfromLDR} by swapping $\hat{\rho}$ and $\hat{\sigma}$.

By using Eq.~\eqref{eq:distratio1} and Eq.~\eqref{eq:QDHfromLDR}, we find that
\begin{equation}
\label{eq:distratio_QDH2}
\inf_{\substack{\rho\in\dop{\hh}\\\rank{\rho}\leq \min\{r,\frac{\dim\hh}{2}\}}}\mu_{\rho,\mathbb{M}}(1)=\min_{\substack{\hat{\rho},\hat{\sigma}\in\dop{\hh},\tr{\hat{\rho}\hat{\sigma}}=0\\\rank{\hat{\rho}}\leq\min\{r,\frac{\dim\hh}{2}\}}}\frac{1}{2}\lMnorm{\hat{\rho}-\hat{\sigma}}\leq
\inf_{\substack{\rho\in\dop{\hh},\sigma\neq\rho\\\rank{\rho}\leq r}}\frac{\lMnorm{\rho-\sigma}}{\lpnorm{1}{\rho-\sigma}}=\inf_{\substack{\rho\in\dop{\hh}\\\rank{\rho}\leq r}}\hat{\mu}_{\rho,\mathbb{M}}.
\end{equation}
Note that we can take the minimum (rather than the infimum) of $\frac{1}{2}\lMnorm{\hat{\rho}-\hat{\sigma}}$ in Eq.~\eqref{eq:distratio_QDH2} due to the continuity of the semi-norm and the compactness of the region of $(\hat{\rho},\hat{\sigma})$.
This guarantees that there exists an optimal $\hat{\rho}$ that satisfies $\rank{\hat{\rho}}\leq \min\{r,\frac{\dim\hh}{2}\}$ and minimizes $\mu_{\rho,\mathbb{M}}(1)$ in Eq.~\eqref{eq:distratio_QDH2}. Since $\hat{\mu}_{\hat{\rho},\mathbb{M}}\leq \mu_{\hat{\rho},\mathbb{M}}(1)$, we can verify that $\hat{\rho}$ also minimizes $\hat{\mu}_{\rho,\mathbb{M}}$ in Eq.~\eqref{eq:distratio_QDH2}.
The middle part of Eq.~\eqref{eq:distratio_QDH} can be obtained from the following calculation.
\begin{equation}
\min_{\substack{\rho\in\dop{\hh}\\\rank{\rho}\leq r}}\hat{\mu}_{\rho,\mathbb{M}}\leq \min_{\substack{\rho\in\dop{\hh}\\\rank{\rho}\leq r}}\mu_{\rho,\mathbb{M}}(1)\leq\min_{\substack{\rho\in\dop{\hh}\\\rank{\rho}\leq \min\{r,\frac{\dim\hh}{2}\}}}\mu_{\rho,\mathbb{M}}(1).
\end{equation}
\end{proof}

We can verify the second statement on the duality between QSV and QDH mentioned in the introduction as follows: If the verification of a subspace $\vv$ has high sample complexity for some QSV parameter $\epsilon$, we find that $\gamma_{\vv,\mathbb{M}}(\epsilon)$ is a small number. By using Theorem \ref{thm:comparison}, this implies $\mu_{\vv,\mathbb{M}}(\epsilon)\leq \gamma_{\vv,\mathbb{M}}(\epsilon)$. 
Moreover by the definition of $\mu_{\vv,\mathbb{M}}(\epsilon)$, there exists a pair $(\rho,\sigma)$ of states such that $\range{\rho}\subseteq\vv$, $\frac{1}{2}\lpnorm{1}{\rho-\sigma}\geq\epsilon$, and $\frac{1}{2}\lMnorm{\rho-\sigma}=\mu_{\vv,\mathbb{M}}(\epsilon)\epsilon$.
Thus, Theorem \ref{thm:constructionQDHfromQSV} implies that there exists a pair $(\hat{\rho},\hat{\sigma})$ of orthogonal states such that $\frac{1}{2}\lMnorm{\hat{\rho}-\hat{\sigma}}\leq\mu_{\vv,\mathbb{M}}(\epsilon)$ and $\rank{\hat{\rho}}\leq\min\{\dim\vv,\frac{\dim\hh}{2}\}$.
This further implies that $(\hat{\rho},\hat{\sigma})$ is $(1,\gamma_{\vv,\mathbb{M}}(\epsilon))$-data hiding.
Note that identifying a pair $(\rho,\sigma)$ might be computationally intractable due to the complexity of the underlying optimization problem. In contrast, it is possible to construct a pair $(\hat{\rho},\hat{\sigma})$ from $(\rho,\sigma)$ simply by using matrix diagonalization, as in the proof of Theorem \ref{thm:constructionQDHfromQSV}.

Fourth, we show that there is a close relationship between extremal subspaces (minimizing visibility) and states (minimizing distinguishability ratio) and that the visitability and distinguishability ratios coincide for such extremal instances. This theorem proves the third statement on the duality between QSV and QDH mentioned in the introduction.
\begin{theorem}
\label{thm:equivalence}
    For any natural number $r<\dim\hh$ and a measurement class $\mathbb{M}$ that is informationally complete, we define a constant $\mu_{r,\mathbb{M}}:=\min_{\rho:\rank{\rho}\leq r}\hat{\mu}_{\rho,\mathbb{M}}$, representing the minimum distinguishability ratio over quantum states with restricted rank. Then, the minimum spectral gap for subspaces with restricted dimension satisfies
    \begin{equation}
 \min_{\vv\subseteq\hh:\dim\vv\leq r}\hat{\gamma}_{\vv,\mathbb{M}}=\mu_{r,\mathbb{M}}.
\end{equation}
Moreover, for any $\hat{\vv}_r=\arg\min_{\vv\subseteq\hh:\dim\vv\leq r}\gamma_{\vv,\mathbb{M}}(1)$ and any $\hat{\rho}_r=\arg\min_{\rho:\rank{\rho}\leq r}\mu_{\rho,\mathbb{M}}(1)$, there exists a state $\rho$ such that $\range{\rho}\subseteq \hat{\vv}_r$ and
    \begin{eqnarray}
        \forall \epsilon\in(0,1], \gamma_{\hat{\vv}_r,\mathbb{M}}(\epsilon)=\mu_{\hat{\rho}_r,\mathbb{M}}(\epsilon)=\gamma_{\range{\hat{\rho}_r},\mathbb{M}}(\epsilon)= \mu_{\rho,\mathbb{M}}(\epsilon)=\mu_{r,\mathbb{M}}.
\end{eqnarray}
\end{theorem}
\begin{proof}
Theorem \ref{thm:constructionQDHfromQSV} implies that there exists a state $\hat{\rho}_r$ such that $\rank{\hat{\rho}_r}\leq r$ and $\mu_{\hat{\rho}_r,\mathbb{M}}(1)=\mu_{r,\mathbb{M}}$. Since $\mu_{\rho,\mathbb{M}}(\epsilon)$ is monotonically non-decreasing, we find that $\mu_{\rho,\mathbb{M}}(1)\geq \hat{\mu}_{\rho,\mathbb{M}}\geq\mu_{r,\mathbb{M}}$ for any state $\rho$ with rank at most $r$ by applying Proposition \ref{prop:limit_distratio}. This implies
\begin{equation}
\label{eq:r1min}
 \min_{\rho:\rank{\rho}\leq r}\mu_{\rho,\mathbb{M}}(1)=\mu_{r,\mathbb{M}}.
\end{equation}
Thus, for any $\hat{\rho}_r=\arg\min_{\rho:\rank{\rho}\leq r}\mu_{\rho,\mathbb{M}}(1)$,
\begin{equation}
\label{eq:repsmin}
  \forall \epsilon\in(0,1], \mu_{\hat{\rho}_r,\mathbb{M}}(\epsilon)=\mu_{r,\mathbb{M}}
\end{equation}
holds due to Proposition \ref{prop:limit_distratio} and the monotonically non-decreasing property.
By using Eq.~\eqref{eq:minimaxform}, we also obtain that
\begin{eqnarray}
 \gamma_{\range{\hat{\rho}_r},\mathbb{M}}(1)&=&\min_{\substack{\rho\in\dop{\hh}:\range{\rho}\subseteq\range{\hat{\rho}_r}\\\sigma\in\dop{\hh}:\tr{\sigma\Pi_\range{\hat{\rho}_r}}=0}}\frac{1}{2}\lMnorm{\rho-\sigma}\\
 \label{eq:upper_visibility1}
 &\leq& \mu_{\hat{\rho}_r,\mathbb{M}}(1)=\mu_{r,\mathbb{M}},
\end{eqnarray}
where we have used Eq.~\eqref{eq:distratio1} to derive the inequality.
On the other hand, Theorem \ref{thm:comparison} implies that
\begin{eqnarray}
\label{eq:lower_visibility1}
  \hat{\gamma}_{\vv,\mathbb{M}}&\geq&\min_{\rho\in\dop{\hh}:\rank{\rho}\leq r}\hat{\mu}_{\rho,\mathbb{M}}=\mu_{r,\mathbb{M}}
\end{eqnarray}
for any subspace $\vv$ with dimension at most $r$.
Eq.~\eqref{eq:upper_visibility1} and Eq.~\eqref{eq:lower_visibility1} together imply
\begin{equation}
  \forall \epsilon\in(0,1],\gamma_{\range{\hat{\rho}_r},\mathbb{M}}(\epsilon)=\hat{\gamma}_{\range{\hat{\rho}_r},\mathbb{M}}=\mu_{r,\mathbb{M}},
\end{equation}
where we have used the monotonically non-decreasing property of $\gamma_{\vv,\mathbb{M}}$ and Proposition \ref{prop:QSV}.

Moreover, Eq.~\eqref{eq:upper_visibility1}, Eq.~\eqref{eq:lower_visibility1}, Proposition \ref{prop:increasing_visibility} and Proposition \ref{prop:QSV} imply that
\begin{equation}
\min_{\vv\subseteq\hh:\dim\vv\leq r}\gamma_{\vv,\mathbb{M}}(1)=\mu_{r,\mathbb{M}}.
\end{equation}
Thus, from Proposition \ref{prop:QSV} and the monotonically non-decreasing property, for any $\hat{\vv}_r=\arg\min_{\vv\subseteq\hh:\dim\vv\leq r}\gamma_{\vv,\mathbb{M}}(1)$, Eq.~\eqref{eq:lower_visibility1} implies that
\begin{equation}
  \forall \epsilon\in(0,1], \gamma_{\hat{\vv}_r,\mathbb{M}}(\epsilon)=\mu_{r,\mathbb{M}}.
\end{equation}

Moreover, Theorem \ref{thm:comparison} implies that
\begin{eqnarray}
   \min_{\rho:\range{\rho}\subseteq\hat{\vv}_r}\mu_{\rho,\mathbb{M}}(1)\leq  \gamma_{\hat{\vv}_r,\mathbb{M}}(1)=\mu_{r,\mathbb{M}}.
\end{eqnarray}
By letting $\rho=\arg\min_{\rho:\range{\rho}\subseteq\hat{\vv}_r}\mu_{\rho,\mathbb{M}}(1)$, Eq.~\eqref{eq:r1min} implies that
$ \mu_{\rho,\mathbb{M}}(1)=\mu_{r,\mathbb{M}}$. Since $\rho=\arg\min_{\rho:\rank{\rho}\leq r}\mu_{\rho,\mathbb{M}}(1)$, Eq.~\eqref{eq:repsmin} implies that
\begin{equation}
  \forall \epsilon\in(0,1],  \mu_{\rho,\mathbb{M}}(\epsilon)=\mu_{r,\mathbb{M}}.
\end{equation}

\end{proof}

The consequence of this theorem is illustrated in Fig.~\ref{fig:diagram}.

\section{Application}
Here, we utilize the duality between QSV and QDH that we established above to show bidirectional implications from QDH to QSV and vice versa.

\subsection{Implications from QDH to QSV}
Here, we show the existence or an explicit construction of QSV protocols and explore their fundamental limitations by using the knowledge of QDH. Note that, from Proposition \ref{prop:samplecomplexity}, the QSV protocols can be implemented by performing identical binary measurements on each sample of $\rho$.

\subsubsection{QSV based on non-adaptive single-qudit measurements}
 Lancien and Winter \cite{LW13} derived the following inequality:
 For any $n$-qudit states $\rho$ and $\sigma$, it holds that
\begin{equation}
    \lpnorm{\mathbb{LO}_n}{\rho-\sigma}\geq\frac{1}{\sqrt{18^n}}\lpnorm{2}{\rho-\sigma}\geq\frac{1}{2\sqrt{18^n\rank{\rho}}}\lpnorm{1}{\rho-\sigma},
\end{equation}
where $\mathbb{LO}_n$ is a measurement realizable by $n$-partite non-adaptive LOCC. Here, we use the same argument to derive Eq.~\eqref{eq:2LOratio}.
By additionally invoking Theorem \ref{thm:comparison}, we obtain that
\begin{equation}
\label{eq:lowerboundLOn}
\hat{\gamma}_{\vv,\mathbb{LO}_n}\geq\inf_{\rho:\rank{\rho}\leq \dim\vv} \hat{\mu}_{\rho,\mathbb{LO}_n}\geq\frac{1}{2\sqrt{18^n\dim\vv}}.
\end{equation}

For the case $\vv=\vee_n\cd$, we can explicitly construct a QSV protocol based on a universal strategy that provides a better lower bound on the spectral gap compared with the general bound, Eq.~\eqref{eq:lowerboundLOn}.
The strategy consists three steps: (1) perform a Haar random unitary transformation on every qudit; (2) perform a measurement with respect to the computational basis $\{\ket{x}\}_{x\in\{0,\cdots,d-1\}^n}$; (3) check whether all the measurement outcomes coincide, i.e., $x_1=\cdots=x_n$.
We can calculate the strategy $\Omega$ as follows.
\begin{eqnarray}
    \Omega=\sum_{i=0}^{d-1}\int(u\ketbra{i}u^\dag)^{\otimes n}du=\frac{d}{\dim\vee_n\cd}\Pi_{\vee_n\cd}.
\end{eqnarray}
This implies that $\hat{\gamma}_{\vee_n\cdim{d},\mathbb{LO}_n}\geq\frac{d}{\dim\vee_n\cd}$.
In particular, since $\dim\vee_n\cd=\begin{pmatrix}
    n+d-1\\d-1
\end{pmatrix}$, we find that $\hat{\gamma}_{\vee_n\cdim{2},\mathbb{LO}_n}\geq\frac{2}{n+1}$.
Moreover, observe that the following strategy
\begin{equation}
    \Omega_k:=\sum_{x\in\{0,1\}^n,x_1+\cdots+x_n=k}\ketbra{x}
\end{equation}
is implementable with a computational basis measurement, and probabilistic choice of measurements enables us to implement the following strategy:
\begin{eqnarray}
    &&(1-p)\Omega+\frac{p}{|K|}\sum_{k\in K}\Omega_k\\
    &=&\frac{2(1-p)}{n+1}\sum_{k=0}^n\ketbra{D_k^{(n)}}+\frac{p}{|K|}\sum_{k\in K}\Omega_k\\
    &=&\left(\frac{2(1-p)}{n+1}+\frac{p}{|K|}\right)\sum_{k\in K}\ketbra{D_k^{(n)}}+\frac{p}{|K|}\sum_{k\in K}(\Omega_k-\ketbra{D_k^{(n)}})+\frac{2(1-p)}{n+1}\sum_{k\notin K}\ketbra{D_k^{(n)}},
\end{eqnarray}
where $\ket{D_k^{(n)}}=\begin{pmatrix}
    n\\k
\end{pmatrix}^{-\frac{1}{2}}\sum_{x\in\{0,1\}^n,x_1+\cdots+x_n=k}\ket{x}$
is the Dicke state and $K\subseteq\{0,1,\cdots,n\}$. Since $(\Omega_k-\ketbra{D_k^{(n)}})$ is a projector, we find that the eigenspace associated with the largest eigenvalue is $\vv:=\vspan{\{\ket{D_k^{(n)}}\}_{k\in K}}$ and the spectral gap between the largest and second largest eigenvalue is $\min\left\{\frac{2(1-p)}{n+1},\frac{p}{|K|}\right\}$.
By setting $p=\frac{2|K|}{n+2|K|+1}$, the spectral gap becomes $\frac{2}{n+2|K|+1}$. Therefore, $\hat{\gamma}_{\vv,\mathbb{LO}_n}\geq\frac{2}{n+2|K|+1}\geq\frac{2}{3(n+1)}$.

\subsubsection{QSV based on PPT measurements}
 Lancien and Winter \cite{LW13} also derived the following inequality:
 For any $n$-qudit states $\rho$ and $\sigma$, it holds that
\begin{equation}
    \lpnorm{\mathbb{PPT}_n}{\rho-\sigma}\geq\lpnorm{2}{\rho-\sigma}\geq\frac{1}{2\sqrt{\rank{\rho}}}\lpnorm{1}{\rho-\sigma},
\end{equation}
where $\mathbb{PPT}_n$ is a measurement realizable by $n$-partite PPT POVMs. Here, we use the same argument to derive Eq.~\eqref{eq:2LOratio}.
By invoking with Theorem \ref{thm:comparison}, we obtain that
\begin{equation}
\label{eq:lowerboundPPTn}
\hat{\gamma}_{\vv,\mathbb{PPT}_n}\geq\inf_{\rho:\rank{\rho}\leq \dim\vv} \hat{\mu}_{\rho,\mathbb{PPT}_n}\geq\frac{1}{2\sqrt{\dim\vv}}.
\end{equation}

For the case $\dim\vv=1$, we can explicitly construct a QSV protocol based on a universal strategy that provides a better lower bound on the spectral gap compared with the general bound, Eq.~\eqref{eq:lowerboundPPTn}.
We define a universal strategy for a target state $\ket{\Phi}\in(\cd)^{\otimes n}$:
\begin{equation}
    \Omega_\Phi=\Phi+\frac{1}{3}(\idop-\Phi).
\end{equation}
We can verify that this strategy is an element of a PPT POVM as follows.
For every subset $K\subseteq\{1,2,\cdots,n\}$, we perform a Schmidt decomposition $\ket{\Phi}=\sum_{i}\sqrt{s_i}\ket{x_i}\ket{y_i}$, where $s_1\geq s_2\geq\cdots$ and $\{\ket{x_i}\}_i$ and $\{\ket{y_i}\}_i$ are sets of orthonormal vectors in qudits in $K$ and their complement $K^c$, respectively. (When $|K|\in\{0,n\}$, we regard that $s_1=1$.) Since any eigenvalue $\lambda$ of the partial transpose $\Phi^{T_K}$ with respect to the cut between $K$ and $K^c$ satisfies $-\sqrt{s_1s_2}\leq\lambda\leq s_1$ \cite{YDY14}, we find that
\begin{eqnarray}
    \Omega_\Phi^{T_K}&=&\frac{2}{3}\Phi^{T_K}+\frac{1}{3}\idop\geq\frac{1-2\sqrt{s_1s_2}}{3}\idop\geq0\\
    (\idop-\Omega_\Phi)^{T_K}&=&\frac{2}{3}\left(\idop-\Phi^{T_K}\right)\geq\frac{2(1-s_1)}{3}\idop\geq0,
\end{eqnarray}
where we use the conditions $s_1,s_2\in[0,1]$ and $s_1+s_2\leq1$.

Therefore, we obtain $\hat{\gamma}_{\Phi,\mathbb{PPT}_n}\geq\min_{\sigma:\tr{\Phi\sigma}=0}\tr{\Omega_\Phi(\Phi-\sigma)}=\frac{2}{3}$ for any target state $\ket{\Phi}$. Moreover, we can show
\begin{equation}
\label{eq:univconstPPT}
    \min_{\Phi}\hat{\gamma}_{\Phi,\mathbb{PPT}_n}=\frac{2}{3}
\end{equation}
by using the following example of $\ket{\Phi}$ attaining the minimum.
Let a target state be a $n$-qudit pure state $\ket{\Phi}$ defined by
\begin{equation}
    \ket{\Phi}:=\ket{\phi^+}^{(12)}\ket{\psi}^{(3\cdots n)},
\end{equation}
where $\ket{\phi^+}=\frac{1}{\sqrt{2}}(\ket{00}+\ket{11})\in\cd\otimes\cd$. Then, any universal strategy $\Omega$ for verifying $\ket{\Phi}$ can be represented as
\begin{equation}
    \Omega=a\Phi+bM,
\end{equation}
where $\tr{\Phi M}=0$, $M\geq0$, $\lpnorm{\infty}{M}=1$, and $0\leq a,b\leq 1$.
Since we have assumed that $\Omega$ is a PPT operator, we find that
\begin{eqnarray}
    0&\leq&(\bra{\phi^-}\bra{\psi})\Omega^{T_1}(\ket{\phi^-}\ket{\psi})\\
    &=&-\frac{a}{2}+b(\bra{\phi^-}\bra{\psi})M^{T_1}(\ket{\phi^-}\ket{\psi})\\
    &=&-\frac{a}{2}+b\tr{((\phi^-)^{T_1}\otimes\psi)M}\\
    &=&-\frac{a}{2}+b\tr{\left(\left((\phi^-)^{T_1}+\frac{1}{2}\phi^+\right)\otimes\psi\right)M}\\
    &=&-\frac{a}{2}+\frac{b}{2}\tr{((\idop_2\otimes\idop_2-\phi^+)\otimes\psi)M}\\
    &\leq&-\frac{a}{2}+\frac{b}{2}\tr{(\idop_2\otimes\idop_2-\phi^+)\otimes\psi}    
    =-\frac{a}{2}+\frac{3b}{2},
\end{eqnarray}
where $\ket{\phi^-}=\frac{1}{\sqrt{2}}(\ket{01}-\ket{10})$, $\idop_2=\ketbra{0}+\ketbra{1}$, and we have used $\bra{\phi^-}(\phi^+)^{T_1}\ket{\phi^-}=-\frac{1}{2}$, $\tr{\Phi M}=0$, $(\phi^-)^{T_1}+\phi^+=\frac{1}{2}\idop_2\otimes\idop_2$, and $M\leq\idop$.
This implies $b\geq\frac{a}{3}$. Since the spectral gap of $\Omega$ is $a-b(\leq\frac{2}{3}a\leq\frac{2}{3})$, we obtain
\begin{equation}
    \hat{\gamma}_{\Phi,\mathbb{PPT}_n}\leq\frac{2}{3},
\end{equation}
which implies Eq.~\eqref{eq:univconstPPT}.

For the case $\vv=\vee_n\cd$, we show upper bounds on $\gamma_{\vee_n\cd,\mathbb{PPT}_n}(1)$.
First, an upper bound for the specific case ($n=2$) can be derived as follows.
\begin{eqnarray}
    \gamma_{\vee_2\cd,\mathbb{PPT}}(1)&\leq&\frac{1}{2}\lpnorm{\mathbb{PPT}}{\beta\Pi_{\vee_2\cd}-\alpha\Pi_{\wedge_2\cd}}
    \leq\frac{2}{d},
\end{eqnarray}
where we have used Proposition \ref{prop:visibilityminimax} and Eq.~\eqref{eq:SymPPTbound}.

Second, a general upper bound can be derived by using the following result of Harrow:

\noindent\textbf{Theorem 7.}\cite{H23}
For any states $\rho,\sigma\in\dop{(\cd)^{\otimes n}}$ that commute with all $U^{\otimes n}$, it holds that
\begin{equation}
    \lpnorm{\mathbb{PPT}_n}{\rho-\sigma}\leq\frac{12n^2}{\sqrt{d}}.
\end{equation}

By using this result, we obtain
\begin{eqnarray}
    \gamma_{\vee_n\cd,\mathbb{PPT}_n}(1)&\leq&\frac{1}{2}\lpnorm{\mathbb{PPT}_n}{\begin{pmatrix}n+d-1\\n\end{pmatrix}^{-1}\Pi_{\vee_n\cd}-\left(d^n-\begin{pmatrix}n+d-1\\n\end{pmatrix}\right)^{-1}(\idop-\Pi_{\vee_n\cd})}\leq\frac{6n^2}{\sqrt{d}},
\end{eqnarray}
where we have used Proposition \ref{prop:visibilityminimax} and the fact that $\Pi_{\vee_n\cd}(U^{\otimes n})=(U^{\otimes n})\Pi_{\vee_n\cd}$ for all $U$.

\subsubsection{QSV based on 4-design POVMs}
First, let us recall the definition of $t$-design POVMs:
\begin{definition}
For a natural number $t$, a finite set $\{\tilde{\phi}_i\}_{i\in I}\subset\pos{\cd}$ of unnormalized pure states is called a $t$-design POVM if it satisfies
\begin{equation}
 \sum_{i\in I}\tilde{\phi}_i^{\otimes t}=d\int \phi^{\otimes t}d\phi =\frac{d}{
\begin{pmatrix}
 t+d-1\\t
\end{pmatrix}
}\Pi_{\vee_t\cd},
\end{equation}
where $\vee_t\cd$ is a symmetric subspace in $(\cd)^{\otimes t}$.
\end{definition}
We can verify that $\{\tilde{\phi}_i\}_{i\in I}$ is a resolution of unity by observing $\sum_{i\in I}\tilde{\phi}_i=\ptr{2,3,\cdots,t}{d\int \phi^{\otimes t}d\phi}=\idop$.
We denote the set of POVMs realized by a $t$-design POVM with classical pre- and post-processing by $\mathbb{DES}_4$.
Note that the binary measurement associated with a $t$-design POVM is formally defined as $\mathbf{DES}_4:=\conv{\left\{\sum_{i\in I'}\tilde{\phi}_i:I'\subseteq I\right\}}$.

We will use the following results derived by several authors~\cite{AE07,MSA09}:
For quantum states $\rho$ and $\sigma$, it holds that
\begin{equation}
 \lpnorm{\mathbb{DES}_4}{\rho-\sigma}\geq\frac{1}{3}\lpnorm{2}{\rho-\sigma}
 \geq\frac{1}{6\sqrt{\rank{\rho}}}\lpnorm{1}{\rho-\sigma},
\end{equation}
where the last inequality follows from the same argument used to derive Eq.~\eqref{eq:2LOratio}.
By additionally invoking Theorem \ref{thm:comparison}, we obtain that
\begin{equation}
\label{eq:lowerboundDES}
\hat{\gamma}_{\vv,\mathbb{DES}_4}\geq\inf_{\rho:\rank{\rho}\leq \dim\vv} \hat{\mu}_{\rho,\mathbb{DES}_4}\geq\frac{1}{6\sqrt{\dim\vv}}.
\end{equation}
Note that we used the fact that $t(\geq2)$-design POVMs are informationally complete \cite{D14}.

\subsection{Implications from QSV to QDH}
Gupta et al.~\cite[Theorem 1]{GHO25} have recently shown that
\begin{equation}
 \forall\ket{\Phi}\in(\cdim{2})^{\otimes n} ,  \hat{\gamma}_{\Phi,\mathbb{LOCC}_n}\geq\frac{1}{n},
\end{equation}
where $\mathbb{LOCC}_n$ represents the measurement class of adaptive single-qubit measurements.
Thus, Theorem \ref{thm:equivalence} implies
\begin{equation}
    \forall\ket{\Phi}\in(\cdim{2})^{\otimes n},\hat{\mu}_{\Phi,\mathbb{LOCC}_n}\geq\frac{1}{n}.
\end{equation}

Gupta et al.~\cite[Theorem 2]{GHO25} have also shown that
there exists a sequence of $n$-qubit pure states $\{\ket{\psi_n}\in(\cdim{2})^{\otimes n}\}_n$ and mixed states $\{\sigma_n\in\dop{(\cdim{2})^{\otimes n}}\}_n$ such that
\begin{equation}
    \lpnorm{\mathbb{LPV}_n}{\psi_n-\sigma_n}=2^{-\Omega(n)},\ \frac{1}{2}\lpnorm{1}{\psi_n-\sigma_n}=1-2^{-\Omega(n)},
\end{equation}
where $\mathbb{LPV}_n$ represents the class of non-adaptive single-qubit projective measurements.
By using Theorem \ref{thm:equivalence}, we obtain
\begin{equation}
    \min_{\ket{\Phi}\in(\cdim{2})^{\otimes n}}\mu_{\Phi,\mathbb{LPV}_n}(1)\leq\hat{\mu}_{\psi_n,\mathbb{LPV}_n}=2^{-\Omega(n)}.
\end{equation}

\section{Discussion}
While QDH was originally proposed within the local measurement scenario, similar phenomena have been studied for other classes of measurements, such as $t$-design measurements \cite{MSA09}, stabilizer measurements \cite{KZG16}, and Gaussian measurements \cite{KA21,YG25}. Researchers have also extensively explored the use of restricted measurements in applications of quantum learning \cite{CCHL21,CGY24} and pseudo-random quantum states \cite{J18}. The findings of QDH and state discrimination presented in those studies will continue to foster exploration on these rapidly growing topics.
Our work enables one to leverage these extensive bodies of research to design QSV protocols across a wide range of subspaces and measurement classes, thereby broadening the applicability and scope of QSV.

Recent breakthroughs in QSV have shown that the power of local measurements—adaptive versus non-adaptive, LOCC versus PPT—can dramatically affect sample complexity, underscoring the sensitivity of local distinguishability in the $n$-partite and asymptotic ($n \rightarrow \infty$) regimes to measurement capabilities. This sensitivity has been little explored in QDH, largely because far fewer multipartite protocols exist compared with bipartite ones, yet it is expected to be even more pronounced for the verification of large-dimensional subspaces. Our work contributes to a deeper understanding of these phenomena, enriching the study of QDH and the broader theory of quantum state discrimination.

\section*{Acknowledgments}
S.A. was partially supported by JST PRESTO Grant no.JPMJPR2111, JST Moonshot R\&D MILLENNIA Program (Grant no.JPMJMS2061), JPMXS0120319794, and CREST (Japan Science and Technology Agency) Grant no.JPMJCR2113.
Y.T. was partially supported by the MEXT Quantum Leap Flagship Program (MEXT Q-LEAP) Grant Number JPMXS0120319794.

\bibliographystyle{plain}

\appendix
\section{$\epsilon$-visibility and sample complexity for QSV}
\label{appendix:samplecomplexity}
\begin{proposition}
\label{prop:samplecomplexity}
 For $\epsilon\in(0,1)$, a nontrivial proper subspace $(\{0\}\subsetneq)\vv(\subsetneq\hh)$, and a measurement class $\mathbb{M}$, the sample complexity $\mathtt{\#sample}(\vv,\mathbb{M},\epsilon,\delta)$ of QSV that decide whether a given state $\rho$ satisfies $\range{\rho}\subseteq\vv$ or $\tr{\rho\Pi_\vv}\leq 1-\epsilon$ with probability at least $1-\delta$ with sequential measurements consisting of ones in $\mathbb{M}$ is bounded as
 \begin{equation}
 \frac{1}{\gamma_{\vv,\mathbb{M}}(\epsilon)}\left(\log\left(\frac{1}{1-\epsilon}\right)\right)^{-1}\log\left(\frac{1}{2\delta}\right)
 \leq \mathtt{\#sample}(\vv,\mathbb{M},\epsilon,\delta)\leq\left\lfloor\frac{2}{ \gamma_{\vv,\mathbb{M}}(\epsilon)^2\epsilon^2}\log\left(\frac{1}{\delta}\right)\right\rfloor+1
\end{equation}
 where  $\gamma_{\vv,\mathbb{M}}(\epsilon)$ is the $\epsilon$-visibility. Furthermore, a QSV protocol satisfying these bounds can be implemented by conducting identical measurements with a binary outcome $\{0,1\}$ on each sample of $\rho$ and then simply adding the outcomes together.
\end{proposition}
\begin{proof}
 From the definition of $\gamma_{\vv,\mathbb{M}}(\epsilon)$, for any real number $s<\gamma_{\vv,\mathbb{M}}(\epsilon)$, there exists a binary POVM $\{\hat{\Omega},\idop-\hat{\Omega}\}$ satisfying $\hat{\Omega}\in\mathbf{M}$ and
 \begin{equation}
\min_{\substack{\rho\in\dop{\hh}:\range{\rho}\subseteq\vv\\\sigma\in\dop{\hh}:\tr{\sigma\Pi_\vv}\leq1-\epsilon}}\tr{\hat{\Omega}(\rho-\sigma)}=\alpha-\beta> \epsilon s,
\end{equation}
where $\alpha:=\min_{\rho:\range{\rho}\subseteq\vv}\tr{\hat{\Omega}\rho}$ and $\beta:=\max_{\sigma:\tr{\sigma\Pi_\vv}\leq1-\epsilon}\tr{\hat{\Omega}\sigma}$.
Suppose a QSV protocol that performs this POVM to each $\rho$ and accepts $\rho$ if the ratio of obtaining the masurement outcome corresponding to $\hat{\Omega}$ is greater than $\frac{\alpha+\beta}{2}$. By using the Hoeffding bound, we find that measuring $m=\lceil2\log(\delta^{-1})(s\epsilon)^{-2}\rceil$ copies of $\rho$ is sufficient for the QSV described in the proposition.
Since $\lim_{x\rightarrow \hat{x}+}\lceil x\rceil=\lfloor\hat{x}\rfloor+1$, this provides the upper bound in the proposition.

To show the lower bound, we let the $i$th measurement be $\{M^{(i)}_{x_i|x_{i-1}\cdots x_1}\}_{x_i}\in\mathbb{M}$ conditioned on the previous measurement outcomes $x_{i-1},\cdots,x_2,x_1$. Since we determine whether to accept or reject $\rho$ based on all the measurement outcomes, the total variation distance of the probability distribution on the measurement outcomes must satisfy
\begin{equation}
\label{eq:lb1sample}
 \frac{1}{2} \sum_{x_m,\cdots,x_1}\left|p\left(x_m,\cdots,x_1|\rho\right)-p\left(x_m,\cdots,x_1|\sigma\right)\right|\geq1-2\delta
\end{equation}
for any states $\rho$ and $\sigma$ such that $\range{\rho}\subseteq\vv$ and $\tr{\sigma\Pi_\vv}\leq1-\epsilon$, where the probability distribution is given by
\begin{eqnarray}
 p(x_1|\rho)&=&\tr{M^{(1)}_{x_1}\rho},\\
 p\left(x_i|x_{i-1}\cdots,x_1,\rho\right) &=& \tr{M^{(i)}_{x_i|x_{i-1}\cdots x_1}\rho},\\
 p\left(x_m,\cdots,x_1|\rho\right) &=& \prod_{i=2}^m p\left(x_i|x_{i-1}\cdots,x_1,\rho\right) p(x_1|\rho).
\end{eqnarray}

By using the identity $1-\frac{1}{2}\sum_x|p(x)-q(x)|=\sum_x\min\{p(x),q(x)\}$ for probability distributions $p(x)$ and $q(x)$, we obtain the following bound.
\begin{eqnarray}
&& 1- \frac{1}{2} \sum_{x_m,\cdots,x_1}\left|p\left(x_m,\cdots,x_1|\rho\right)-p\left(x_m,\cdots,x_1|\sigma\right)\right|\\
&=&\sum_{x_m,\cdots,x_1}\min\left\{p\left(x_m,\cdots,x_1|\rho\right),p\left(x_m,\cdots,x_1|\sigma\right)\right\}\\
&=&\sum_{x_m,\cdots,x_1}\min\{p\left(x_m|x_{m-1},\cdots,x_1,\rho\right)p\left(x_{m-1},\cdots,x_1|\rho\right),
p\left(x_m|x_{m-1},\cdots,x_1,\sigma\right)p\left(x_{m-1},\cdots,x_1|\sigma\right)\}\\
&\geq&\sum_{x_m,\cdots,x_1}\min\{p\left(x_m|x_{m-1},\cdots,x_1,\rho\right),p\left(x_m|x_{m-1},\cdots,x_1,\sigma\right)\}\min\{p\left(x_{m-1},\cdots,x_1|\rho\right),p\left(x_{m-1},\cdots,x_1|\sigma\right)\}\nonumber\\\\
&\geq&\sum_{x_m,\cdots,x_1}\prod_{i=2}^m\min\{p\left(x_i|x_{i-1},\cdots,x_1,\rho\right),p\left(x_i|x_{i-1},\cdots,x_1,\sigma\right)\} \min\{p\left(x_1|\rho\right),p\left(x_1|\sigma\right)\}.
\label{eq:lb2sample}
\end{eqnarray}

From Proposition \ref{prop:visibilityminimax}, there exist states $\hat{\rho}$ and $\hat{\sigma}$ such that $\range{\hat{\rho}}\subseteq\vv$, $\tr{\hat{\sigma}\Pi_\vv}\leq1-\epsilon$, and $\frac{1}{2}\lMnorm{\hat{\rho}-\hat{\sigma}}=\epsilon \gamma_{\vv,\mathbb{M}}(\epsilon)$.
Thus, we obtain 
\begin{eqnarray}
\sum_{x_1}\min\{p(x_1|\hat{\rho}),p(x_1|\hat{\sigma})\}&=&1- \frac{1}{2}\sum_{x_1}|p(x_1|\hat{\rho})-p(x_1|\hat{\sigma})|\\
&\geq&1-\frac{1}{2}\lMnorm{\hat{\rho}-\hat{\sigma}}=1-\epsilon \gamma_{\vv,\mathbb{M}}(\epsilon),\\
\sum_{x_i}\min\{p\left(x_i|x_{i-1},\cdots,x_1,\hat{\rho}\right),p\left(x_i|x_{i-1},\cdots,x_1,\hat{\sigma}\right)\} &=&1- \frac{1}{2}\sum_{x_i}|p\left(x_i|x_{i-1},\cdots,x_1,\hat{\rho}\right)-p\left(x_i|x_{i-1},\cdots,x_1,\hat{\sigma}\right)|\nonumber\\\\
&\geq&1-\frac{1}{2}\lMnorm{\hat{\rho}-\hat{\sigma}}=1-\epsilon \gamma_{\vv,\mathbb{M}}(\epsilon)
\end{eqnarray}
for any conditioning variables $x_{i-1},\cdots,x_1$ and any $i\in\{2,3,\cdots,m\}$.
Since $\hat{\rho}$ and $\hat{\sigma}$ satisfy Eq.~\eqref{eq:lb1sample}, by combining with Eq.~\eqref{eq:lb2sample}, we obtain
\begin{equation}
 2\delta\geq\left(1-\epsilon \gamma_{\vv,\mathbb{M}}(\epsilon)\right)^m.
\end{equation}
This implies $m\geq\left(\log\frac{1}{1-\epsilon \gamma_{\vv,\mathbb{M}}(\epsilon)}\right)^{-1}\log\frac{1}{2\delta}$.
Since $f(x)=\log\frac{1}{1-x}$ is a convex function in $x\in[0,1)$, $\forall s\in[0,1],\log\frac{1}{1-\epsilon s}=f(s\epsilon+(1-s)0)\leq sf(\epsilon)+(1-s)f(0)=s\log\frac{1}{1-\epsilon}$.
This completes the proof since Proposition \ref{prop:increasing_visibility} and $\gamma_{\vv,\mathbb{M}}(1)\leq1$ together guarantee $\gamma_{\vv,\mathbb{M}}\in[0,1]$.
\end{proof}

\section{Existence of ball around $\frac{1}{2}\idop$}
\label{appendix:eball}
In this section, we prove a lemma regarding the existence of a ball around $\frac{1}{2}\idop$ in a binary measurement $\mathbf{M}$ that is associated with a measurement class $\mathbb{M}$ that is informationally complete.

\begin{lemma}
\label{lemma:ball}
 For a binary measurement $\mathbf{M}$ associated with a measurement class $\mathbb{M}$ that is informationally complete, i.e., the real span of $\mathbf{M}$ is equal to the set of Hermitian operators, there exists a positive number $r(>0)$ such that $\frac{1}{2}\idop+H\in\mathbf{M}$ for any  Hermitian operator $H$ such that $\lpnorm{\infty}{H}\leq r$.
\end{lemma}
\begin{proof}
 Since $\mathbf{M}$ spans the set of Hermitian operators, there exists a basis $\{M_i\in\mathbf{M}\}_i$.
 For Hermitian operator $H$, we define $l_1$-norm $|H|_1$ with respect to this basis, i.e., $|H|_1:=\sum_i |r_i|$ where $H=\sum_i r_iM_i$. 
It is known that any two norms are equivalent in finite-dimensional vector spaces. In particular, there exists a positive number $r(>0)$ such that $\lpnorm{\infty}{H}\leq r$ implies $|H|_1\leq\frac{1}{2}$. Suppose a Hermitian operator $H$ satisfies $|H|_1\leq\frac{1}{2}$ and can be decomposed as $H=\sum_i r_iM_i$.
We can show $\frac{1}{2}\idop+H\in\mathbf{M}$ by using the identity
\begin{eqnarray}
\label{eq:decomposition}
\frac{1}{2}\idop+H&=&\left(\frac{1}{2}-\sum_{i\in R_-}|r_i|\right)\idop+\sum_{i\in R_+}r_iM_i+\sum_{i\in R_-}|r_i|(\idop-M_i),
\end{eqnarray}
where $R_\pm=\{i:\pm r_i>0\}$. This implies $\frac{1}{2}\idop+H\in\mathbf{M}$ because $\frac{1}{2}-\sum_{i\in R_-}|r_i|\geq\frac{1}{2}-|H|_1\geq0$, $\left(\frac{1}{2}-\sum_{i\in R_-}|r_i|\right)+\sum_{i\in R_+}r_i+\sum_{i\in R_-}|r_i|\leq\frac{1}{2}+|H|_1\leq1$, $\{0,\idop,M_i,(\idop-M_i)\}\subseteq\mathbf{M}$ and $\mathbf{M}$ is convex.
\end{proof}

\section{Continuity of $\epsilon$-visibility}
\label{appendix:continuity}
\begin{proposition}
\label{prop:continuity_visibility}
 For a nontrivial proper subspace $(\{0\}\subsetneq)\vv(\subsetneq\hh)$ and a measurement class $\mathbb{M}$, $\epsilon$-visibility $\gamma_{\vv,\mathbb{M}}(\epsilon)$ is continuous in $\epsilon\in(0,1]$.
\end{proposition}
\begin{proof}
 Since $\gamma_{\vv,\mathbb{M}}(\epsilon)$ is monotonically non-decreasing, it is sufficient to show the following continuity bound:
 \begin{equation}
 \label{eq:continuitybound}
\forall \epsilon,\epsilon'\ s.t.\  0<\epsilon<\epsilon'\leq1, \gamma_{\vv,\mathbb{M}}(\epsilon')\leq \gamma_{\vv,\mathbb{M}}(\epsilon)+f(\epsilon,\epsilon'),
\end{equation}
where $\lim_{\epsilon\rightarrow\epsilon'-}f(\epsilon,\epsilon')=\lim_{\epsilon'\rightarrow\epsilon+}f(\epsilon,\epsilon')=0$.
From Proposition \ref{prop:visibilityminimax}, we can let $\gamma_{\vv,\mathbb{M}}(\epsilon)=\frac{1}{2\epsilon}\lMnorm{\hat{\rho}-\hat{\sigma}}$, where $\range{\hat{\rho}}\subseteq\vv$ and $\tr{\hat{\sigma}\Pi_\vv}\leq1-\epsilon$. Since $1-\epsilon<1$, we can define a state $\hat{\sigma}_\bot=\frac{1}{\tr{\Pi_{\vv_\bot}\hat{\sigma}}}\Pi_{\vv_\bot}\hat{\sigma}\Pi_{\vv_\bot}$ that is orthogonal to $\Pi_\vv$. Futhermore, by defining a state
\begin{equation}
 \sigma'=(1-p)\hat{\sigma}+p\hat{\sigma}_\bot=\hat{\sigma}+p(\hat{\sigma}_\bot-\hat{\sigma}),
\end{equation}
where $1-p=\frac{1-\epsilon'}{1-\epsilon}$, we can verify that $\tr{\sigma'\Pi_\vv}=(1-p)\tr{\hat{\sigma}\Pi_\vv}\leq1-\epsilon'$. Thus, we find that
\begin{eqnarray}
 \gamma_{\vv,\mathbb{M}}(\epsilon')&\leq&\frac{1}{2\epsilon'}\lMnorm{\hat{\rho}-\sigma'}\\
 &\leq&\frac{1}{2\epsilon'}\lMnorm{\hat{\rho}-\hat{\sigma}}+\frac{p}{2\epsilon'}\lMnorm{\hat{\sigma}_\bot-\hat{\sigma}}\\
 &=&\frac{\epsilon}{\epsilon'}\gamma_{\vv,\mathbb{M}}(\epsilon)+\frac{\epsilon'-\epsilon}{1-\epsilon}\frac{\lMnorm{\hat{\sigma}_\bot-\hat{\sigma}}}{2\epsilon'}\\
 &\leq&\gamma_{\vv,\mathbb{M}}(\epsilon)+\frac{\epsilon'-\epsilon}{1-\epsilon}\frac{\lpnorm{1}{\hat{\sigma}_\bot-\hat{\sigma}}}{2\epsilon'},
\end{eqnarray}
where we have used $\frac{\epsilon}{\epsilon'}\leq1$ to derive the last inequality. 
We proceed with the calculation as follows.
\begin{eqnarray}
 \lpnorm{1}{\hat{\sigma}_\bot-\hat{\sigma}}&\leq&\lpnorm{1}{\Pi_\vv\hat{\sigma}\Pi_\vv}+2\lpnorm{1}{\Pi_\vv\hat{\sigma}\Pi_{\vv_\bot}}+\left(\frac{1}{\tr{\Pi_{\vv_\bot}\hat{\sigma}}}-1\right)\lpnorm{1}{\Pi_{\vv_\bot}\hat{\sigma}\Pi_{\vv_\bot}}\\
 &=&2\left(\tr{\Pi_\vv\hat{\sigma}}+\lpnorm{1}{\Pi_\vv\hat{\sigma}\Pi_{\vv_\bot}}\right)\\
 &\leq&2\left(\tr{\Pi_\vv\hat{\sigma}}+\lpnorm{2}{\Pi_\vv\sqrt{\hat{\sigma}}}\lpnorm{2}{\sqrt{\hat{\sigma}}\Pi_{\vv_\bot}}\right)\\
 &=&2\left(\tr{\Pi_\vv\hat{\sigma}}+\sqrt{\tr{\Pi_\vv\hat{\sigma}}}\sqrt{\tr{\Pi_{\vv_\bot}\hat{\sigma}}}\right)\\
 &\leq&2\left(1-\epsilon+\sqrt{1-\epsilon}\right),
\end{eqnarray}
where we have used H\"older's inequality for unitarily invariant norms \cite[Corollary IV.2.6]{RBBook} to derive the second inequality.
Therefore, we obtain a continuity bound in the form of Eq.~\eqref{eq:continuitybound} with $f(\epsilon,\epsilon')=\frac{1-\epsilon+\sqrt{1-\epsilon}}{\epsilon'}\frac{\epsilon'-\epsilon}{1-\epsilon}$. We can verify $\lim_{\epsilon\rightarrow\epsilon'-}f(\epsilon,\epsilon')=\lim_{\epsilon'\rightarrow\epsilon+}f(\epsilon,\epsilon')=0$ for any $\epsilon\in(0,1)$ and $\epsilon'\in(0,1]$.
\end{proof}

\end{document}